\documentclass[reprint,aps,prb,twocolumn,amsmath,amssymb,showpacs,preprintnumbers]{revtex4-2}

\usepackage{graphicx}[]
\usepackage{dcolumn}
\usepackage{bm}
\usepackage{multirow}

\usepackage{braket}
\usepackage{color}
\usepackage{ulem}
\usepackage{url}
\usepackage{hyperref}

\begin{document}

\title{
Chern insulating state with double-$Q$ ordering wave vectors \\ 
at the Brillouin zone boundary
}
\author{Satoru Hayami}
\affiliation{
Graduate School of Science, Hokkaido University, Sapporo 060-0810, Japan
}

\begin{abstract}
Magnetic multiple-$Q$ states consisting of multiple spin density waves are a source of unconventional topological spin textures, such as skyrmion and hedgehog.
We theoretically investigate a topologically nontrivial double-$Q$ state with a net spin scalar chirality on a two-dimensional square lattice. 
We find that a double-$Q$ spiral superposition of the ordering wave vectors located at the Brillouin zone boundary gives rise to unconventional noncoplanar spin textures distinct from the skyrmion crystal. 
We show that such a double-$Q$ state is stabilized by the interplay among the easy-axis anisotropic interaction, high-harmonic wave-vector interaction, and external magnetic field. 
Furthermore, the obtained double-$Q$ state becomes a Chern insulating state with a quantum Hall conductivity when the Fermi level is located in the band gaps. 
Our present results provide another platform to realize topological magnetic states other than skyrmion crystals by focusing on the symmetry of constituent ordering wave vectors in momentum space. 
\end{abstract}

\maketitle

\section{Introduction}

A Chern insulator is one of the insulators in two-dimensional magnetic systems without time-reversal symmetry~\cite{Haldane_PhysRevLett.61.2015}. 
It is characterized by a topological invariant that is termed a Chern number, which is related to the Hall conductivity $\sigma_{xy}$; $\sigma_{xy}$ is quantized in the unit of $e^2/h$ when the system is insulating ($e$ is the elementary charge and $h$ is the Planck constant)~\cite{Thouless_PhysRevLett.49.405, kohmoto1985topological}. 
The Chern number corresponds to the sum of the Berry fluxes of all the two-dimensional plaquettes for the closed torus surface in momentum space. 
The appearance of such Berry fluxes makes the electron feel an auxiliary magnetic field so that the wave function acquires the spin Berry phase~\cite{berry1984quantal, Loss_PhysRevB.45.13544, Ye_PhysRevLett.83.3737}. 
Especially, noncoplanar spin textures with a spin scalar chirality, which is defined by a triple scalar product of spins, $\bm{S}_i \cdot (\bm{S}_j \times \bm{S}_k)$, can be the origin of Berry fluxes even if neither a net magnetization nor relativistic spin--orbit coupling is present~\cite{Ohgushi_PhysRevB.62.R6065, Shindou_PhysRevLett.87.116801, Nagaosa_RevModPhys.82.1539}. 

In solids, noncoplanar spin textures are often described by a superposition of multiple spin density waves referred to as a multiple-$Q$ state~\cite{hayami2021topological}. 
The Chern insulators have been so far proposed in various noncoplanar spin textures under different lattice structures. 
One of the typical examples is a triple-$Q$ tetrahedron state on a triangular lattice, which is stabilized by the ring-exchange interaction~\cite{Momoi_PhysRevLett.79.2081}, perfect nesting~\cite{Martin_PhysRevLett.101.156402, Chern_PhysRevLett.109.156801, Venderbos_PhysRevLett.108.126405}, or partial nesting by $(d-2)$-dimensional connections of the Fermi surfaces~\cite{Akagi_JPSJ.79.083711, Akagi_PhysRevLett.108.096401, Hayami_PhysRevB.90.060402, hayami_PhysRevB.91.075104, Hayami_PhysRevB.94.024424}. 
Similar multiple-$Q$ states based on the nesting mechanism have been found in other lattice structures, such as tetragonal~\cite{Huang_PhysRevB.102.195120}, checkerboard ~\cite{Venderbos_PhysRevLett.109.166405}, honeycomb ~\cite{Jiang_PhysRevLett.114.216402}, and kagome structures~\cite{Barros_PhysRevB.90.245119, Ghosh_PhysRevB.93.024401}. 
In these cases, the multiple-$Q$ ordering wave vectors lie at the high-symmetric points in the Brillouin zone boundary, which indicates short modulation periods of noncoplanar magnetic structures. 
Another example is a magnetic skyrmion crystal (SkX) with a long-period swirling spin texture~\cite{nagaosa2013topological}, which is stabilized by the Dzyaloshinskii-Moriya (DM) interaction~\cite{dzyaloshinsky1958thermodynamic,moriya1960anisotropic, rossler2006spontaneous, Yi_PhysRevB.80.054416, Butenko_PhysRevB.82.052403, Hayami_PhysRevB.105.014408, lin2021skyrmion}, frustrated exchange interaction~\cite{Okubo_PhysRevLett.108.017206, leonov2015multiply,Lin_PhysRevB.93.064430,Hayami_PhysRevB.93.184413, Hayami_PhysRevB.94.174420, Lin_PhysRevLett.120.077202, Hayami_PhysRevB.103.224418}, dipolar exchange interaction~\cite{Utesov_PhysRevB.103.064414, Utesov_PhysRevB.105.054435}, anisotropic exchange interaction~\cite{amoroso2020spontaneous, yambe2021skyrmion, amoroso2021tuning, Hirschberger_10.1088/1367-2630/abdef9, Hayami_PhysRevB.103.054422, Yambe_PhysRevB.106.174437}, (higher-order) Ruderman-Kittel-Kasuya-Yosida interaction (RKKY)~\cite{Ruderman,Kasuya,Yosida1957, Ozawa_PhysRevLett.118.147205, Hayami_PhysRevB.95.224424, Wang_PhysRevLett.124.207201, Hayami_PhysRevB.99.094420, Mitsumoto_PhysRevB.104.184432, hayami2021phase, Mitsumoto_PhysRevB.105.094427, Kobayashi_PhysRevB.106.L140406, Bouaziz_PhysRevLett.128.157206}, orbital frustration~\cite{nomoto2023ab}, electric dipolar interaction~\cite{muto2023theory}, circularly polarized microwave field~\cite{Eto_PhysRevB.104.104425}, and multiple-spin interaction~\cite{heinze2011spontaneous, Yambe_PhysRevB.107.014417}. 
The SkX can become the Chern insulator when the Fermi level lies in the band gap~\cite{Hamamoto_PhysRevB.92.115417, Gobel_PhysRevB.95.094413, Gobel_PhysRevB.96.060406}. 
In contrast to the first example, constituent ordering wave vectors of the SkXs are located inside the Brillouin zone so that longer modulation periods occur. 
In this way, various noncoplanar spin textures can be ubiquitously described by considering multiple-$Q$ states irrespective of magnetic modulation periods.

\begin{figure}[tb!]
\begin{center}
\includegraphics[width=1.0\hsize]{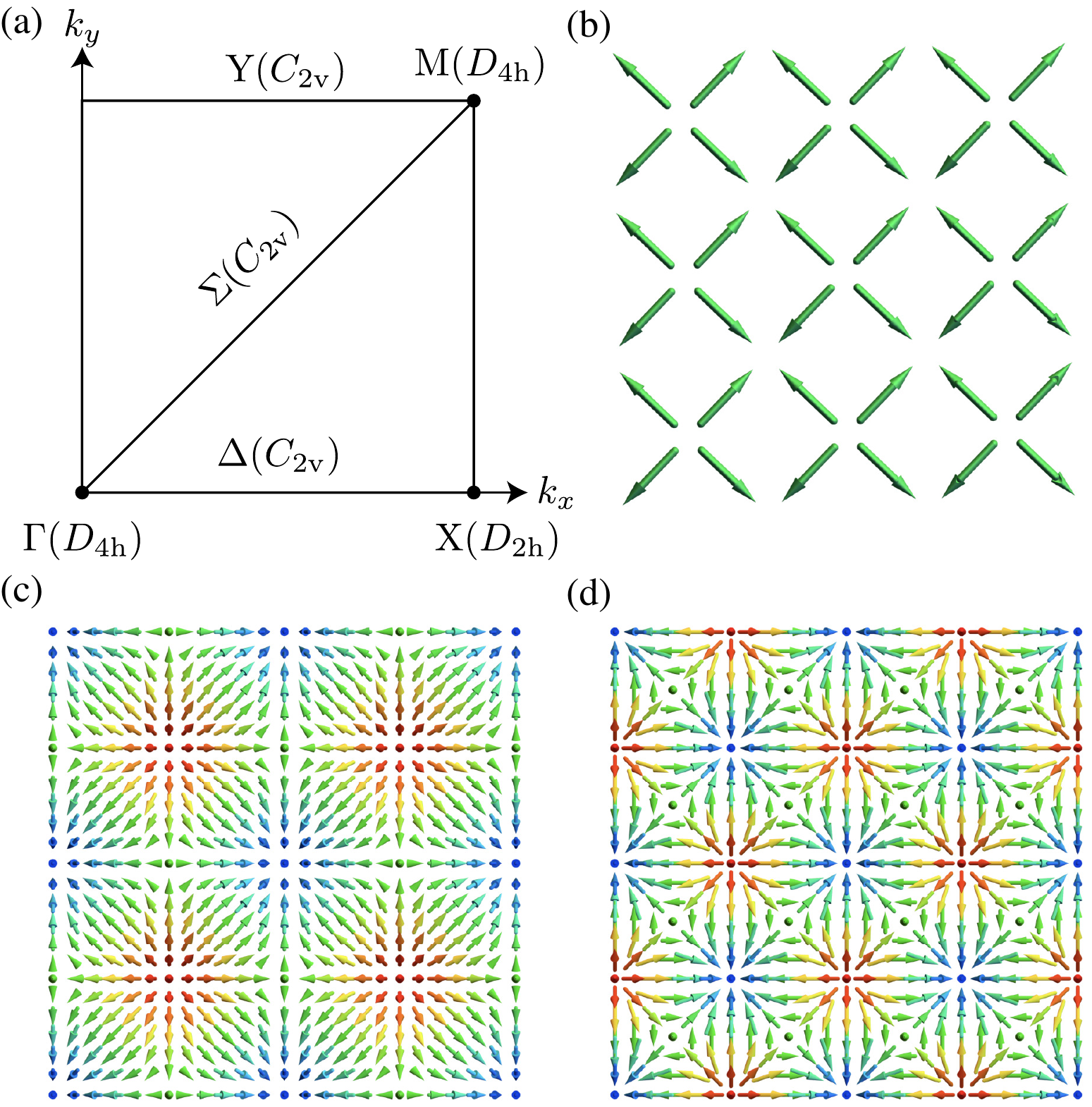} 
\caption{
\label{fig: ponti} 
(a) High-symmetric points and lines of the two-dimensional square lattice in momentum ($k_x$--$k_y$) space under the space group $P4/mmm$: $\Gamma=(0,0)$, ${\rm X}=(\pi, 0)$, ${\rm M}=(\pi,\pi)$, $\Sigma=(u, u)$, $\Delta=(u,0)$, and ${\rm Y}=(u, \pi)$ for $0 < u < \pi$. 
The point-group symmetry in each point and line is also shown. 
(b) The double-$Q$ structure consisting of the ordering wave vectors at the X point. 
(c, d) The square skyrmion crystal consisting of the double-$Q$ ordering wave vectors on the (c) $\Delta$ and (d) $\Sigma$ lines. 
The red, blue, and green arrows represent the positive, negative, and zero $z$ components of the spin moments. 
}
\end{center}
\end{figure}

When considering further possible multiple-$Q$ instability, it is useful to classify the situation based on the symmetry in terms of the constituent ordering wave vectors in momentum space, since multiple-$Q$ states often consist of symmetry-related ordering wave vectors under lattice structures; the double-$Q$ states tend to be favored under the tetragonal lattice structure with fourfold rotational symmetry and the triple-$Q$ states tend to be favored under the hexagonal lattice structure with sixfold rotational symmetry. 
Specifically, we consider a two-dimensional square-lattice system under the space group $P4/mmm$, where the symmetry of the high-symmetric points and lines for the two-dimensional wave vectors $(k_x, k_y)$ in the Brillouin zone is shown in Fig.~\ref{fig: ponti}(a); we set the lattice constant of the square lattice as unity ($a=1$).  
When the ordering wave vectors lie at the $\Gamma=(0,0)$ or ${\rm M}=(\pi,\pi)$ point belonging to the $D_{\rm 4h}$ symmetry, there is no multiple-$Q$ instability owing to no symmetry-related wave vectors in the Brillouin zone. 
Meanwhile, when considering the ${\rm X}=(\pi,0)$ point so that the symmetry of the wave vector is lowered to $D_{\rm 2h}$, the $(0, \pi)$ point corresponds to the symmetry-related wave vectors via the fourfold rotational operation invariant under the square lattice. 
In this case, a coplanar double-$Q$ state with the ordering wave vectors $(\pi,0)$ and $(0, \pi)$ in Fig.~\ref{fig: ponti}(b) can be realized depending on the electronic band structure~\cite{Agterberg_PhysRevB.62.13816, Hayami_PhysRevB.90.060402, Gastiasoro_PhysRevB.92.140506, allred2016double}. 
Moreover, further low-symmetric wave vectors $(u,0)$ and $(0,u)$ [$(u,u)$ and $(-u,u)$] for $0<u<\pi$ on the $\Delta$ ($\Sigma$) line belonging to the $C_{2v}$ symmetry also gives rise to the double-$Q$ SkX in both noncentrosymmetric and centrosymmetric structures, as shown in Fig.~\ref{fig: ponti}(c) [Fig.~\ref{fig: ponti}(d)]: The DM interaction plays an important role for the former~\cite{Hayami_PhysRevLett.121.137202}, while the frustrated exchange interaction~\cite{Wang_PhysRevB.103.104408, Utesov_PhysRevB.103.064414, Hayami_PhysRevB.105.174437}, multiple-spin (many-body)  interaction~\cite{Christensen_PhysRevX.8.041022, Hayami_PhysRevB.103.024439, hayami2022multiple}, the staggered DM interaction~\cite{hayami2022square}, and other effects~\cite{Huang_PhysRevResearch.5.013125} are important for the latter. 
These studies provide a microscopic understanding of the SkX-hosting materials, such as GdRu$_2$Si$_2$~\cite{khanh2020nanometric, Yasui2020imaging, Nomoto_PhysRevLett.125.117204, khanh2022zoology, Bouaziz_PhysRevLett.128.157206, Matsuyama_PhysRevB.107.104421, Wood_PhysRevB.107.L180402} and EuAl$_4$~\cite{Shang_PhysRevB.103.L020405, shimomura2019lattice, kaneko2021charge, Zhu_PhysRevB.105.014423,takagi2022square, Meier_PhysRevB.106.094421, Gen_PhysRevB.107.L020410, hayami2023orthorhombic}.

In the present study, we investigate further topologically nontrivial multiple-$Q$ instability by considering other symmetries of ordering wave vectors in momentum space. 
We here focus on the multiple-$Q$ states with the ordering wave vectors on the ${\rm Y}=(\pi, u)$ and $(u, \pi)$ lines at the Brillouin zone boundary in Fig.~\ref{fig: ponti}(a). 
Although the wave vectors on the Y line have the same symmetry as those on the $\Delta$ and $\Sigma$ lines, we find that a double-$Q$ magnetic chiral (2$Q$ MC) state with a different nontrivial topological spin texture from the conventional square SkX appears in this case. 
Although the obtained spin texture is totally distinct from the SkX, it exhibits a uniform scalar chirality similar to the SkX. 
We show that the synergy among the easy-axis anisotropic interaction, high-harmonic wave-vector interaction, and external magnetic field plays an important role in stabilizing the 2$Q$ MC state. 
Furthermore, this state becomes the Chern insulating state when considering the itinerant electron degree of freedom; the Hall conductivity is quantized as an integer.  
Our results provide a possibility of exotic multiple-$Q$ states by taking into account the symmetry of the constituent ordering wave vectors in momentum space. 

The rest of this paper is organized as follows: 
In Sec.~\ref{sec: Model and method}, we introduce the spin and itinerant electron models on the square lattice and outline the numerical method based on the simulated annealing. 
In Sec.~\ref{sec: Magnetic phase diagram}, we show the ground-state phase diagram of the model and describe the details of the obtained spin and scalar chirality textures. 
In Sec.~\ref{sec: Topologically nontrivial electronic state}, we show that the 2$Q$ MC state corresponds to the magnetic Chern insulator with the quantized Hall conductance once the Fermi level lies in the band gap. 
In Sec.~\ref{sec: Summary}, we present a summary of this paper. 
In Appendix~\ref{appendix}, we show the magnetic phase diagram in the absence of the high-harmonic wave-vector interaction in order to compare the phase diagram in Sec.~\ref{sec: Magnetic phase diagram}. 

\section{Model and method}
\label{sec: Model and method}

\subsection{Localized spin model}

We calculate the ground-state spin configuration by considering an effective spin model with the momentum-resolved interaction~\cite{hayami2021topological}, which is given by 
\begin{align}
\label{eq: Ham}
\mathcal{H}=  &-2J\sum_{\nu=1,2}(\bm{S}_{\bm{Q}_{\nu}} \cdot \bm{S}_{-\bm{Q}_{\nu}}
+ I^{\rm A }S^z_{\bm{Q}_{\nu}} S^z_{-\bm{Q}_{\nu}}) \cr
&-2J'\sum_{\nu=3,4}(\bm{S}_{\bm{Q}_{\nu}} \cdot \bm{S}_{-\bm{Q}_{\nu}}
+ I'^{\rm A }S^z_{\bm{Q}_{\nu}} S^z_{-\bm{Q}_{\nu}}) \cr
&-H \sum_i S^z_i,
\end{align}
where $\bm{S}_{\bm{Q}_{\nu}}$ is the $\bm{Q}_\nu$ component of the spin moment; $\bm{Q}_\nu$ is the ordering wave vectors for $\nu =$1--4 and $\bm{S}_{\bm{Q}_{\nu}}$ is derived from the Fourier transform of the classical localized spin $\bm{S}_i$ with fixed length $|\bm{S}_i|=1$. 
The first term represents the bilinear exchange interaction at the dominant wave-vector channels $\bm{Q}_1$ and $\bm{Q}_2$; we set $\bm{Q}_1=(\pi/4, \pi)$ and $\bm{Q}_2=(\pi, -\pi/4)$ so that they are located on the Y line at the Brillouin zone boundary in Fig.~\ref{fig: ponti}(a). 
It is noted that $\bm{Q}_1$ and $\bm{Q}_2$ are connected by the fourfold rotational symmetry of the square lattice. 
We introduce the easy-axis anisotropic exchange interaction $I^{\rm A}>0$ in order to enhance the multiple-$Q$ instability~\cite{leonov2015multiply, Lin_PhysRevB.93.064430, Hayami_PhysRevB.93.184413}. 
We set $J=1$ as the energy unit of the model. 

The second term represents the bilinear exchange interaction in the different wave-vector channels at $\bm{Q}_3=(3\pi/4, 3\pi/4)$ and $\bm{Q}_4=(-3\pi/4, 3\pi/4)$.  
The choice of $\bm{Q}_3$ and $\bm{Q}_4$ is owing to the relation satisfying $\bm{Q}_3=-\bm{Q}_1+\bm{Q}_2$ and $\bm{Q}_4=\bm{Q}_1+\bm{Q}_2$ except for the difference by reciprocal lattice vector. 
Thus, $\bm{Q}_3$ and $\bm{Q}_4$ are regarded as the high-harmonic wave vectors of $\bm{Q}_1$ and $\bm{Q}_2$. 
It was shown that the interaction at such high-harmonic wave vectors plays an important role in stabilizing the multiple-$Q$ states including the square SkX~\cite{Hayami_PhysRevB.105.174437, hayami2022multiple, hayami2023widely}. 
We set $J'=0.6 J $ and $I'^{\rm A}=I^{\rm A}/2$ in the following analysis. 
We show the result for $J'=0$ in Appendix~\ref{appendix}, where no multiple-$Q$ instability occurs in the phase diagram. 
The third term represents the Zeeman coupling in the presence of an external magnetic field along the $z$ direction. 

The optimal spin configurations of the model in Eq.~(\ref{eq: Ham}) with changing $I^{\rm A}$ and $H$ are obtained by the numerically simulated annealing. 
For a system size with $N=16^2$ under the periodic boundary condition, we perform Monte Carlo simulations while decreasing the temperature in the following manner. 
We start from a random spin configuration at a high temperature $T_0=1.5$ and reduce the temperature at the rate of $\alpha=0.999999$ to the final temperature $T=0.0001$. 
At each temperature, we update all the spins one by one in real space based on the standard single-spin-flip Metropolis algorithm. 
Finally, we perform $10^5$-$10^6$ Monte Carlo sweeps for thermalization and measurements.
We also start the simulations from the spin patterns obtained at low temperatures to determine the phase boundaries between different magnetic states. 

In order to identify each magnetic phase, we compute spin and chirality quantities. 
In the spin sector, we calculate the spin structure factor, which is given by 
\begin{align}
S^{\eta\eta}_s(\bm{q})= \frac{1}{N} \sum_{i,j}S_i^{\eta} S_j^{\eta} e^{i \bm{q}\cdot (\bm{r}_i -\bm{r}_j)}, 
\end{align}
where $\eta=x,y,z$, $\bm{r}_i$ is the position vector at site $i$, and $\bm{q}$ is the wave vector.  
For the in-plane spin component, we use $S^{\perp}_s(\bm{q})=S^{xx}_s(\bm{q})+S^{yy}_s(\bm{q})$. 
We also compute the $\bm{Q}_\nu$ component of magnetic moments as 
\begin{align}
m^\eta_{\bm{Q}_\nu} = \sqrt{\frac{S^{\eta\eta}_s(\bm{Q}_\nu)}{N}}. 
\end{align}
The net magnetization is given by 
\begin{align}
M^\eta = \frac{1}{N} \sum_i S_i^\eta. 
\end{align}
In the chiral sector, the scalar chirality is given by 
\begin{align}
\chi^{\rm sc}&= \frac{1}{2 N} 
\sum_{i}
\sum_{\delta,\delta'= \pm1}
\delta \delta'
 \bar{\bm{S}}_{i} \cdot (\bar{\bm{S}}_{i+\delta\hat{x}} \times \bar{\bm{S}}_{i+\delta'\hat{y}}), 
\end{align}
where $\hat{x}$ ($\hat{y}$) represents a shift by lattice constant in the $x$ ($y$) direction.

\subsection{Itinerant electron model}
\label{sec: Itinerant electron model}

We analyze the band structure by considering the spin-charge coupled Hamiltonian, which is given by 
\begin{align}
\label{eq: HamKLM}
\mathcal{H} = 
-t \sum_{i, j,  \sigma}  c^{\dagger}_{i\sigma}c_{j \sigma}
+J_{\rm K} \sum_{i} \bm{s}_i \cdot \bm{S}_i,  
\end{align}
where $c^{\dagger}_{i\sigma}$ ($c_{i \sigma}$) is a creation (annihilation) operator of an itinerant electron at site $i$ and spin $\sigma$. 
The first term stands for the hopping of itinerant electrons between the nearest-neighbor sites on the square lattice and the second term stands for the onsite exchange coupling between itinerant electron spins $\bm{s}_i=(1/2)\sum_{\sigma, \sigma'}c^{\dagger}_{i\sigma} \bm{\sigma}_{\sigma \sigma'} c_{i \sigma'}$ and localized spins $\bm{S}_i$ with coupling constant $J_{\rm K}$, where $\bm{\sigma}=(\sigma^x,\sigma^y,\sigma^z)$ is the vector of Pauli matrices. 
We substitute the spin configuration obtained by the simulated annealing to the spin model in Eq.~(\ref{eq: Ham}) into $\bm{S}_i$. 
We set $t=1$ as the energy unit of the model in Eq.~(\ref{eq: HamKLM}) and consider the strong-coupling regime by taking $J_{\rm K}=100$.

\section{Magnetic phase diagram}
\label{sec: Magnetic phase diagram}

\begin{figure}[tb!]
\begin{center}
\includegraphics[width=1.0\hsize]{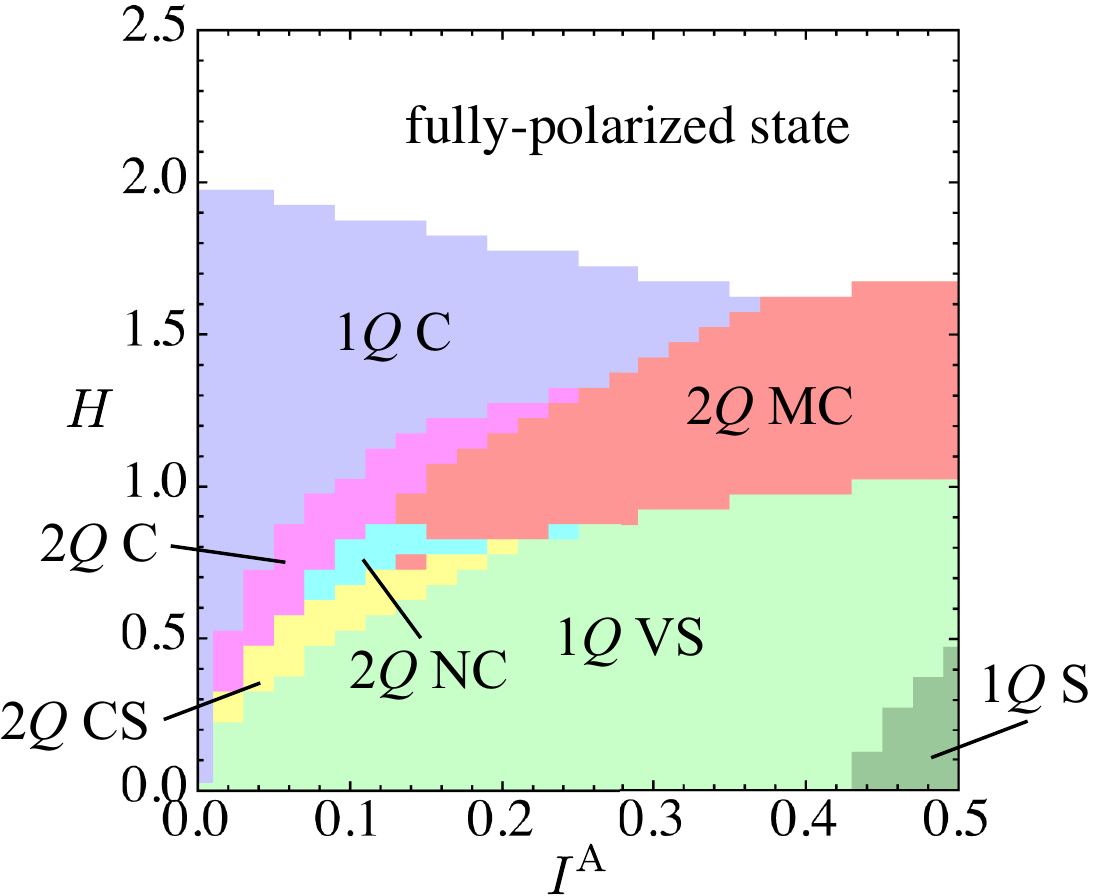} 
\caption{
\label{fig: PD} 
Ground-state phase diagram of the model in Eq.~(\ref{eq: Ham}) on the square lattice, which is obtained by the simulated annealing at $J'=0.6$. 
The horizontal axis represents the easy-axis anisotropic interaction $I^{\rm A}$, while the vertical axis represents the external magnetic field along the $z$ direction. 
1$Q$ and 2$Q$ denote the single-$Q$ and double-$Q$ states, respectively. 
VS, S, C, CS, NC, and MC stand for the vertical spiral, sinusoidal, conical, chiral stripe, nonchiral, and magnetic chiral states, respectively. 
}
\end{center}
\end{figure}

\begin{figure}[tb!]
\begin{center}
\includegraphics[width=1.0\hsize]{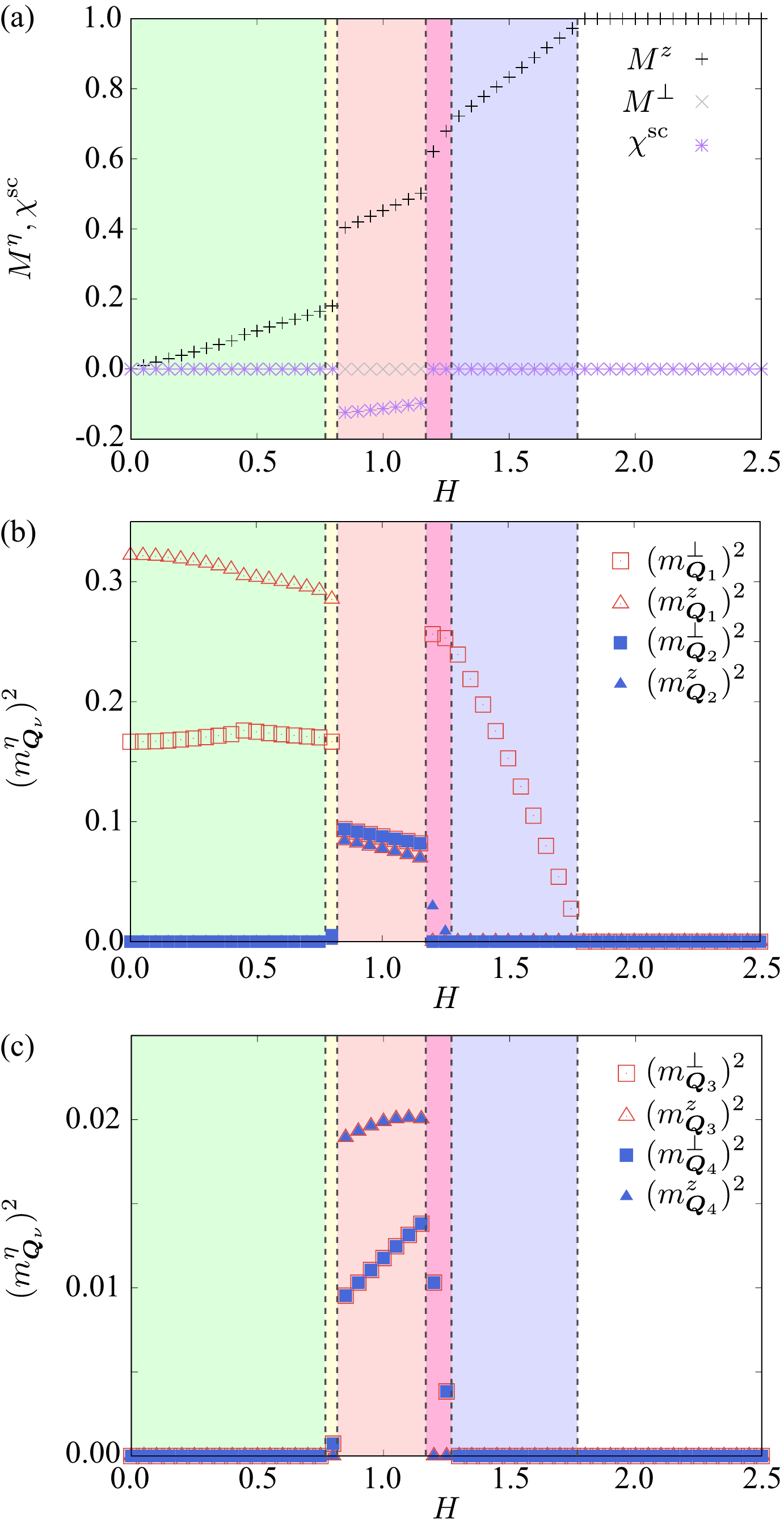} 
\caption{
\label{fig: mag_x=0.2} 
$H$ dependence of (a) the magnetization $M^\eta$ and the scalar chirality $\chi^{\rm sc}$, (b) $\bm{Q}_1$ and $\bm{Q}_2$ components of squared magnetic moments $(m^{\eta}_{\bm{Q}_{1,2}})^2$, and (c) $\bm{Q}_3$ and $\bm{Q}_4$ components of squared magnetic moments $(m^{\eta}_{\bm{Q}_{3,4}})^2$ for $\eta=\perp, z$ at $I^{\rm A}=0.2$. 
The vertical dashed lines represent the phase boundaries between different spin states. 
}
\end{center}
\end{figure}

\begin{figure}[tb!]
\begin{center}
\includegraphics[width=1.0\hsize]{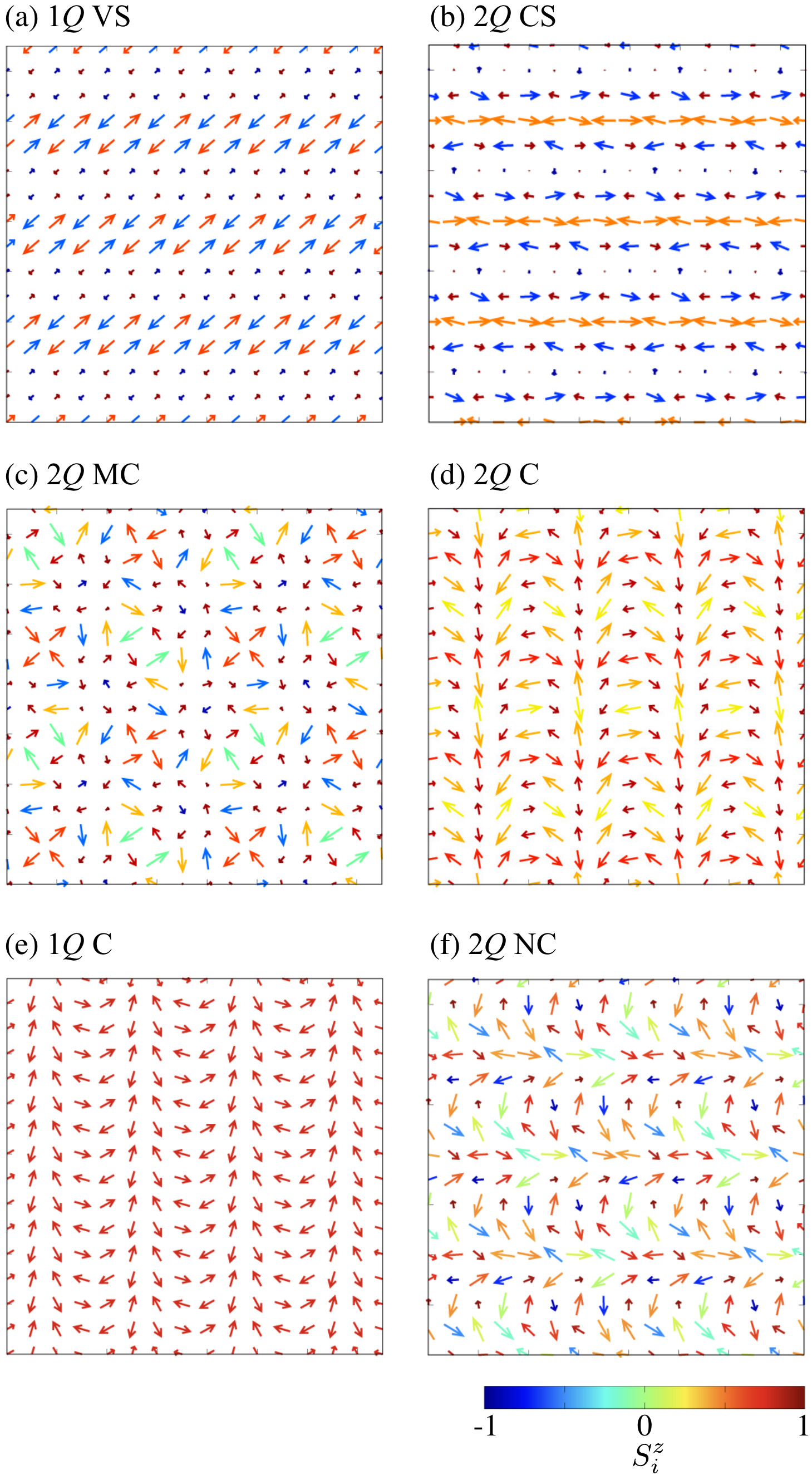} 
\caption{
\label{fig: Spin} 
Real-space spin configurations in 
(a) the 1$Q$ VS state at $I^{\rm A}=0.2$ and $H=0$, 
(b) the 2$Q$ CS state at $I^{\rm A}=0.2$ and $H=0.8$, 
(c) the 2$Q$ MC state at $I^{\rm A}=0.2$ and $H=1$, 
(d) the 2$Q$ C state at $I^{\rm A}=0.2$ and $H=1.2$, 
(e) the 1$Q$ C state at $I^{\rm A}=0.2$ and $H=1.4$, and 
(f) the 2$Q$ NC state at $I^{\rm A}=0.1$ and $H=0.7$. 
The arrows represent the direction of the in-plane spin and the color shows its $z$ component. 
}
\end{center}
\end{figure}

\begin{figure}[tb!]
\begin{center}
\includegraphics[width=0.88\hsize]{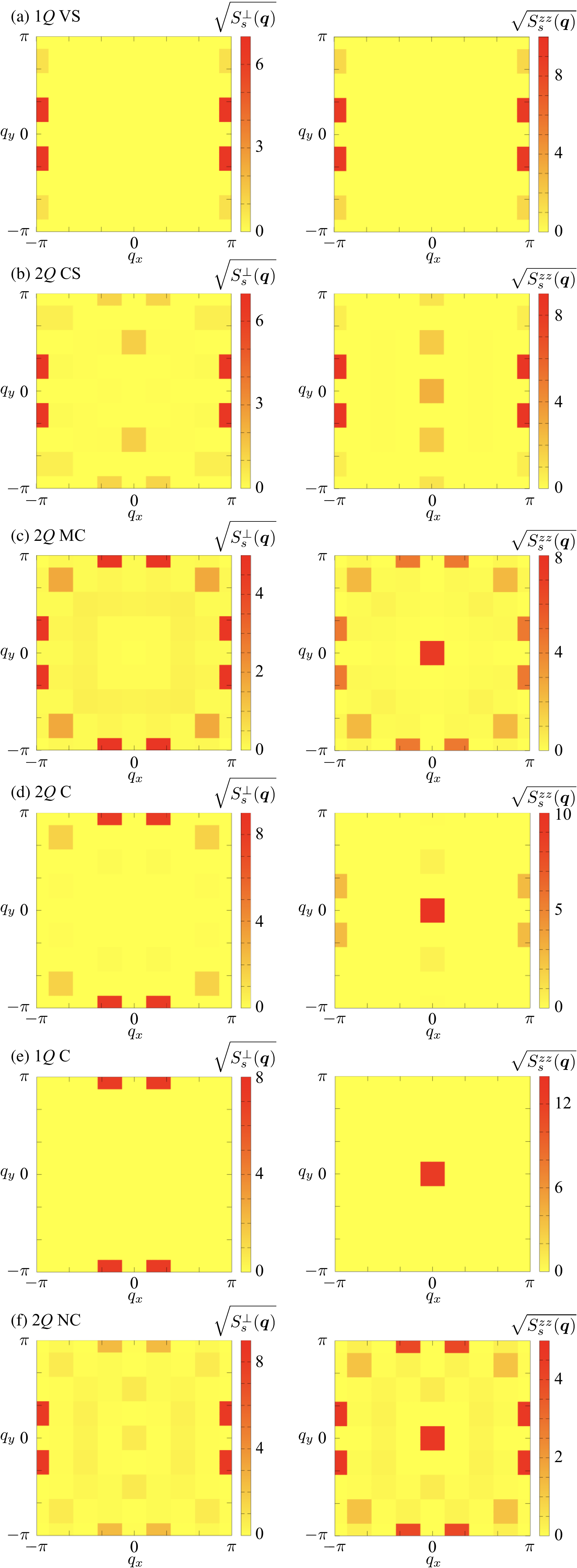} 
\caption{
\label{fig: Sq} 
The square root of the (left panel) $xy$ and (right panel) $z$ components of the spin structure factor in the first Brillouin zone in each phase in Fig.~\ref{fig: Spin}. 
The model parameters are the same as those in Fig.~\ref{fig: Spin}. 
}
\end{center}
\end{figure}

\begin{figure}[tb!]
\begin{center}
\includegraphics[width=1.0\hsize]{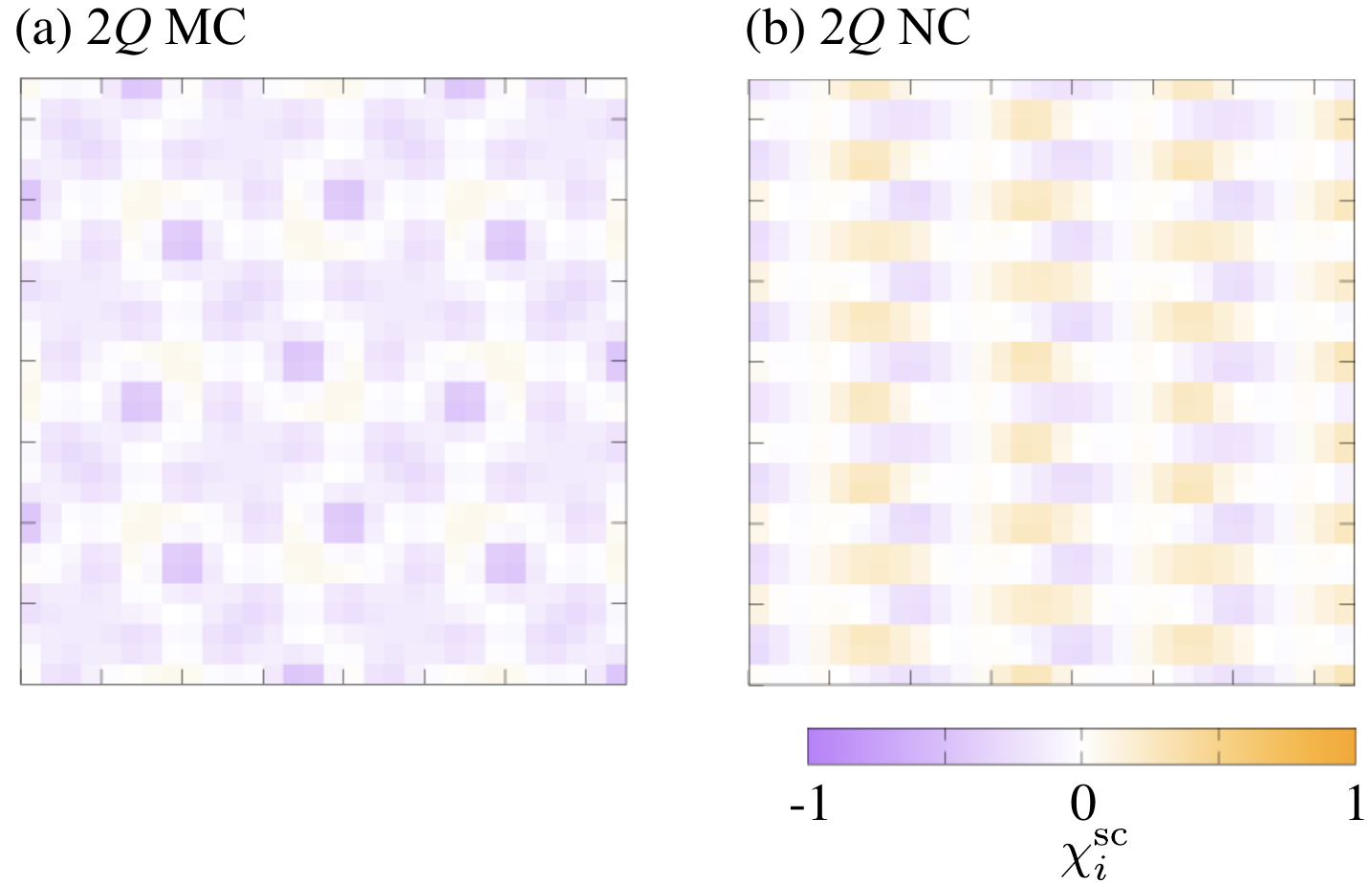} 
\caption{
\label{fig: chirality} 
Real-space scalar chirality configurations in 
(a) the 2$Q$ MC state at $I^{\rm A}=0.2$ and $H=1$ and
(b) the 2$Q$ NC state at $I^{\rm A}=0.1$ and $H=0.7$. 
}
\end{center}
\end{figure}

Figure~\ref{fig: PD} shows the ground-state spin configuration of the spin model in Eq.~(\ref{eq: Ham}) in the plane of $I^{\rm A}$ and $H$. 
The phase diagram includes seven phases in addition to the fully-polarized state with $S_i^z =(0,0,1)$, which appears in the high-field region. 
A phase sequence for $I^{\rm A}=0$ or $H=0$ is simple; the single-$Q$ conical spiral (1$Q$ C) state, whose spiral plane lies on the $xy$ plane, is stabilized for $H \lesssim 2$ when $I^{\rm A}=0$, whereas the single-$Q$ vertical spiral (1$Q$ VS) and single-$Q$ sinusoidal (1$Q$ S) states appear depending on $I^{\rm A}$. 
The spiral plane of the 1$Q$ VS state includes the $z$ direction, where the in-plane direction is arbitrary owing to the absence of the in-plane magnetic anisotropy like the bond-dependent anisotropy. 
The 1$Q$ S state is represented by the single-$Q$ sinusoidal modulation along the $z$ direction.  
When taking into account both $I^{\rm A}$ and $H$, one finds that various double-$Q$ states are stabilized in the phase diagram. 

Among the double-$Q$ states, the 2$Q$ MC state is the only state to have a nonzero net spin scalar chirality, which indicates that this 2$Q$ MC state corresponds to a topologically nontrivial state. 
Thus, we mainly describe the details of the 2$Q$ MC state in the following. 

Figures~\ref{fig: mag_x=0.2}(a)--\ref{fig: mag_x=0.2}(c) show the $H$ dependence of the uniform magnetization $M^\eta$ and the scalar chirality $\chi^{\rm sc}$, $\bm{Q}_1$ and $\bm{Q}_2$ components of squared magnetic moments $(m^{\eta}_{\bm{Q}_{1,2}})^2$, and $\bm{Q}_3$ and $\bm{Q}_4$ components of squared magnetic moments $(m^{\eta}_{\bm{Q}_{3,4}})^2$ for $\eta=\perp, z$ at $I^{\rm A}=0.2$, respectively.
We here show the data in each ordered state by appropriately sorting $(m^\eta_{\bm{Q}_{\nu}})$ for better readability. 
In the low-field region, the 1$Q$ VS state is stabilized, whose spin configuration in real space and the spin structure factor in momentum space are shown in Figs.~\ref{fig: Spin}(a) and \ref{fig: Sq}(a), respectively. 
Since the dominant interaction in the model is the $\bm{Q}_{1,2}$ channel, the 1$Q$ VS state exhibits a single-$Q$ structure at $\bm{Q}_1$ or $\bm{Q}_2$. 
In Fig.~\ref{fig: Spin}(a), the staggered alignment of the spins is found along the $x$ direction in real space owing to the ordering wave vector $\bm{Q}_2=(\pi, -\pi/4)$, which is consistent with the peak structure in momentum space in Fig.~\ref{fig: Sq}(a). 
It is noted that $\bm{Q}_2=(\pi, -\pi/4)$ is identical to $(\pi,\pi/4)$, $(-\pi, \pi/4)$, and $(-\pi, -\pi/4)$ via the translation by the reciprocal lattice vector. 
The small intensity at $3\bm{Q}_2=(\pi, \pm 3\pi/4)$ is owing to the elliptical modulation of the spiral plane that arises from the easy-axis anisotropic interaction $I^{\rm A}$ and/or the magnetic field $H$. 

When $H$ is increased, the 1$Q$ VS state continously changes into the double-$Q$ chiral stripe (2$Q$ CS) state.  
Similarly to the 1$Q$ VS state, the 2$Q$ CS state is mainly characterized by the vertical spiral structure with $\bm{Q}_2$, as shown in Fig.~\ref{fig: mag_x=0.2}(b). 
Meanwhile, in contrast to the 1$Q$ VS state, the $\bm{Q}_1$ component of the magnetic moments, $m^{\perp}_{\bm{Q}_1}$, is slightly induced, as shown in Fig.~\ref{fig: Sq}(b); it is noted that the state with the dominant $\bm{Q}_1$ and subdominant $\bm{Q}_2$ modulations has the same energy [Fig.~\ref{fig: mag_x=0.2}(b)].  
This anisotropic double-$Q$ superposition leads to the noncoplanar spin structure, as shown by the real-space spin configuration in Fig.~\ref{fig: Spin}(b), although the scalar chirality shows a finite-$q$ component rather than the uniform one. 
This is why this state is called the chiral stripe state, whose similar spin textures have been found in itinerant magnets and anisotropic magnets~\cite{Solenov_PhysRevLett.108.096403, Ozawa_doi:10.7566/JPSJ.85.103703, hayami2020multiple}. 
It is noted that the 2$Q$ CS state has the amplitude of $(m^\perp_{\bm{Q}_3})^2=(m^\perp_{\bm{Q}_4})^2$ in Figs.~\ref{fig: mag_x=0.2}(c) and \ref{fig: Sq}(b), which indicates that the high-harmonic wave-vector interaction contributes to their stabilization. 
Indeed, the 2$Q$ CS state vanishes for $J'=0$, as discussed in Appendix~\ref{appendix}. 

Further increment of $H$ in the 2$Q$ CS state drives the phase transition to the 2$Q$ MC state with jumps of $M^z$ and $\chi^{\rm sc}$, as shown in Fig.~\ref{fig: mag_x=0.2}(a). 
The 2$Q$ MC state is characterized by the double-$Q$ structure with the same intensity at $\bm{Q}_1$ and $\bm{Q}_2$ in each spin component, as shown in Figs.~\ref{fig: mag_x=0.2}(b) and \ref{fig: Sq}(c), i.e. $(m^{\perp}_{\bm{Q}_1})^2=(m^{\perp}_{\bm{Q}_2})^2$ and $(m^{z}_{\bm{Q}_1})^2=(m^{z}_{\bm{Q}_2})^2$. 
In addition, this state has the same intensity at $\bm{Q}_3$ and $\bm{Q}_4$, i.e., $(m^{\perp}_{\bm{Q}_3})^2=(m^{\perp}_{\bm{Q}_4})^2$ and $(m^{z}_{\bm{Q}_3})^2=(m^{z}_{\bm{Q}_4})^2$, as shown in Fig.~\ref{fig: mag_x=0.2}(c). 
These features are the same as those in the square SkX~\cite{hayami2023widely}. 
On the other hand, the real-space spin configuration is totally different from the SkX, as shown in Fig.~\ref{fig: Spin}(c); there are no skyrmion cores with $S_i^z=-1$ and the spin variations between the adjacent sites seem to be discrete rather than continuous. 
Nevertheless, this state exhibits uniform scalar chirality, as shown by the real-space scalar chirality configuration in Fig.~\ref{fig: chirality}(a), where the negative net component is found. 
It is noted that the state with the positive scalar chirality has the same energy as that with the negative one owing to the absence of the in-plane magnetic anisotropy, which leads to a degeneracy of the spiral helicity; such a degeneracy can be lifted by considering the in-plane bond-dependent anisotropy for the $\bm{Q}_1$--$\bm{Q}_4$ components, which is allowed under the tetragonal symmetry~\cite{Hayami_doi:10.7566/JPSJ.89.103702, Hayami_PhysRevB.105.104428}. 
Reflecting a nonzero net scalar chirality, the 2$Q$ MC state shows nonzero Berry curvature in the band structure; it becomes the Chern insulator when the Fermi level is located in the band gap, as discussed in Sec.~\ref{sec: Topologically nontrivial electronic state}. 

The emergence of the 2$Q$ MC state is attributed to the interplay between the easy-axis magnetic anisotropy $I^{\rm A}$ and the high-harmonic wave-vector interaction $J'$ in the external magnetic field, which is common to the SkX~\cite{hayami2023widely}; the stability region of the 2$Q$ MC state extends for larger $I^{\rm A}$ and $J'$. 
Their difference is only found in the real-space spin texture that originates from the different positions of the constituent ordering wave vectors in the Brillouin zone; the 2$Q$ MC state is constructed by the double-$Q$ superposition of the ordering wave vectors on the Y line in the Brillouin zone in Fig.~\ref{fig: ponti}, while the SkX is constructed by the superposition of the wave vectors on the $\Delta$ or $\Sigma$ line. 
From these results, one finds that there is a chance of realizing the 2$Q$ MC state by similar microscopic interactions so as to stabilize the SkX, such as the frustrated exchange interaction~\cite{Wang_PhysRevB.103.104408, Utesov_PhysRevB.103.064414, Hayami_PhysRevB.105.174437} and multiple-spin (many-body)  interaction~\cite{Christensen_PhysRevX.8.041022, Hayami_PhysRevB.103.024439, hayami2022multiple}, once the ordering wave vectors lie on the Y line.

\begin{figure}[tb!]
\begin{center}
\includegraphics[width=1.0\hsize]{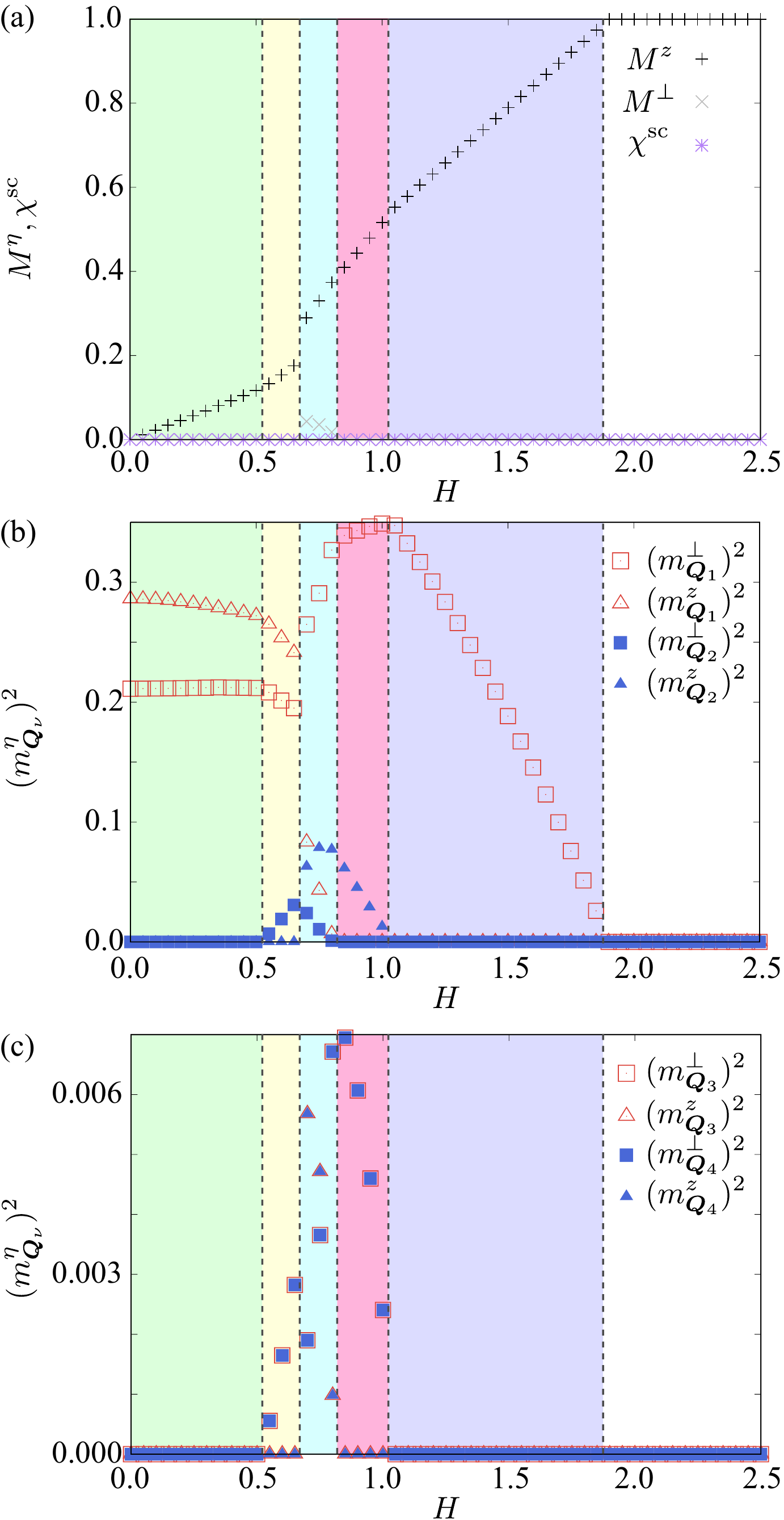} 
\caption{
\label{fig: mag_x=0.1} 
$H$ dependence of (a) $M^\eta$ and $\chi^{\rm sc}$, (b) $(m^{\eta}_{\bm{Q}_{1,2}})^2$, and (c) $(m^{\eta}_{\bm{Q}_{3,4}})^2$ for $\eta=\perp, z$ at $I^{\rm A}=0.1$. 
The vertical dashed lines represent the phase boundaries between different spin states.
}
\end{center}
\end{figure}

The 2$Q$ MC state turns into the double-$Q$ conical (2$Q$ C) state with jumps of $M^z$ and $\chi^{\rm sc}$. 
The 2$Q$ C state is characterized by the conical spiral modulation with the $\bm{Q}_1$ component and the $z$-directional sinusoidal modulation with the $\bm{Q}_2$ component, as shown in Figs.~\ref{fig: mag_x=0.2}(b) and \ref{fig: Sq}(d). 
Similarly to the other double-$Q$ states, there is an intensity of magnetic moments at $\bm{Q}_3$ and $\bm{Q}_4$ in order to gain the energy by $J'$, as shown in Fig.~\ref{fig: mag_x=0.2}(c). 
This state also accompanies the chirality density wave with the $\bm{Q}_2$ component owing to the noncoplanar spin structure arising from the double-$Q$ superposition [Fig.~\ref{fig: Spin}(d)]. 
$(m^\eta_{\bm{Q}_{2,3,4}})^2$ becomes smaller as $H$ increases and vanishes when the phase transition to the 1$Q$ C state occurs. 
The spin configuration and spin structure factor of the 1$Q$ C state are shown in Figs.~\ref{fig: Spin}(e) and \ref{fig: Sq}(e), respectively. 
Finally, the 1$Q$ C state continuously changes into the fully-polarized state.

For small $I^{\rm A}$, the double-$Q$ nonchiral (2$Q$ NC) state appears in the intermediate-field region instead of the 2$Q$ MC state. 
Figures~\ref{fig: mag_x=0.1}(a)--\ref{fig: mag_x=0.1}(c) show the $H$ dependence of magnetization and $\bm{Q}_\nu$ components of the magnetic moments at $I^{\rm A}=0.1$, where the 2$Q$ NC state appears between the 2$Q$ CS and 2$Q$ C states.   
This state is characterized by an anisotropic double-$Q$ state including the high-harmonic wave-vector modulation similar to the 2$Q$ CS and 2$Q$ C states, as shown in Figs.~\ref{fig: Sq}(f), \ref{fig: mag_x=0.1}(b) and \ref{fig: mag_x=0.1}(c). 
The real-space spin texture in Fig.~\ref{fig: Spin}(f) seems to be a complicated noncoplanar spin texture, but there is no net scalar chirality in the magnetic unit cell, as shown in Fig.~\ref{fig: chirality}(b).

\section{Topologically nontrivial electronic state}
\label{sec: Topologically nontrivial electronic state}

We discuss the electronic structure of the 2$Q$ MC state with nonzero scalar chirality by adopting the tight-binding model in Eq.~(\ref{eq: HamKLM}) in Sec.~\ref{sec: Itinerant electron model}. 
First, we discuss the band structure for bulk in the 2$Q$ MC state in Sec.~\ref{sec: Band structure}. 
Then, we present the edge state arising from the nontrivial topological spin texture in Sec.~\ref{sec: Edge state}. 

\subsection{Band structure}
\label{sec: Band structure}

\begin{figure}[tb!]
\begin{center}
\includegraphics[width=1.0\hsize]{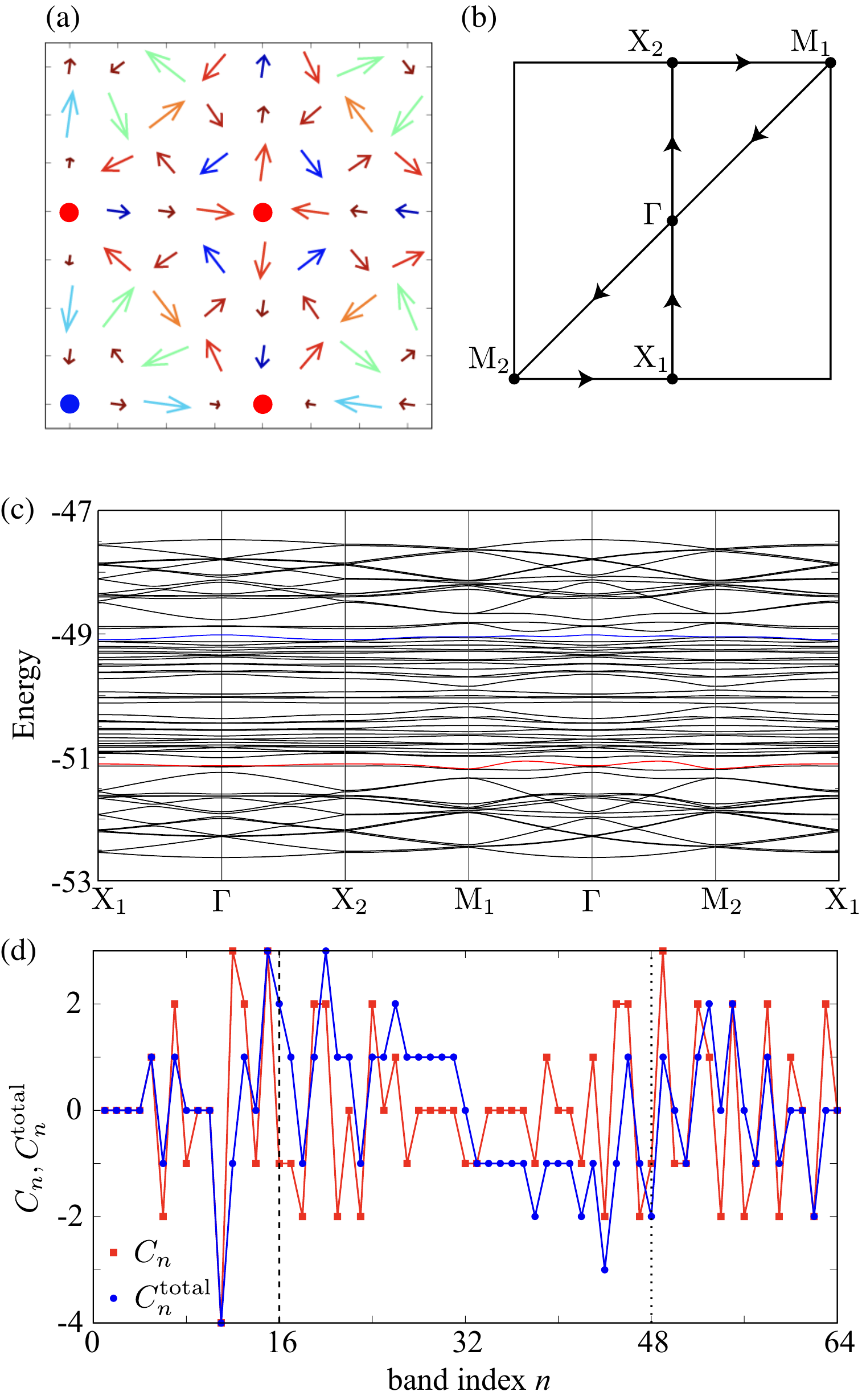} 
\caption{
\label{fig: band} 
(a) The spin configuration of the 2$Q$ MC state in the $8\times 8$ magnetic unit cell at $I^{\rm A}=0.2$, $H=0.85$, and $J'=0.6$. 
The red and blue spheres represent the spin moments with $S^z_i=+1$ and $S^z_i=-1$, respectively. 
(b) The first Brillouin zone of the square lattice. 
(c) The electronic band structure of the spin-charge coupled Hamiltonian in Eq.~(\ref{eq: HamKLM}) in the 2$Q$ MC state at $t=-1$ and $J_{\rm K}=100$ along the high-symmetric lines in (b). 
The bands in red and blue stand for the 16th and 48th bands from the lowest energy band, respectively. 
(d) The Chern number $C_n$ for the $n$th band. The sum of the Chern number up to the $n$th band, $C^{\rm total}_n = \sum^n_m C_m$, is also shown.
The vertical dashed and dotted lines correspond to the 1/8 and 3/8 fillings, respectively. 
}
\end{center}
\end{figure}

\begin{figure}[tb!]
\begin{center}
\includegraphics[width=1.0\hsize]{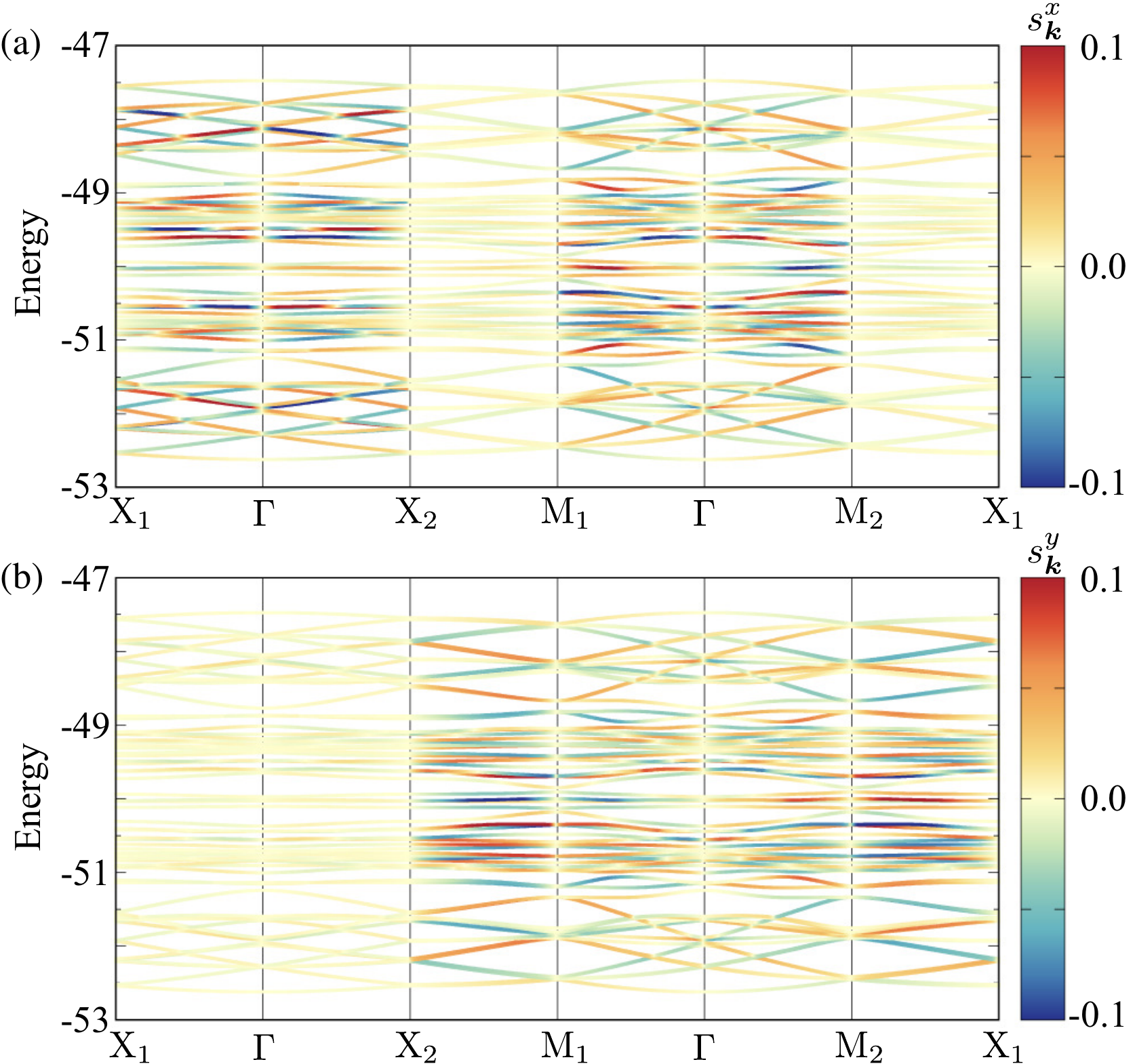} 
\caption{
\label{fig: band_spinsplit} 
Band structure projected onto the (a) $x$ and (b) $y$ components of the momentum-resolved spin, $s^{x,y}_{\bm{k}}$. 
}
\end{center}
\end{figure}

We investigate the electronic band dispersion of the tight-binding model in Eq.~(\ref{eq: HamKLM}) in the presence of the 2$Q$ MC spin texture. 
We use the spin configuration of the 2$Q$ MC state, which is obtained at $I^{\rm A}=0.2$ and $H=0.85$ by the simulated annealing in Sec.~\ref{sec: Magnetic phase diagram}; the real-space spin configuration in the $8\times 8$ magnetic unit cell is shown in Fig.~\ref{fig: band}(a). 

We plot the band structure of the 2$Q$ MC state in Fig.~\ref{fig: band}(c) for the high-symmetric lines in the magnetic Brillouin zone in Fig.~\ref{fig: band}(b). 
Here, we only show 64 out of total 128 bands; the other 64 bands are located around $E \sim 50$ ($E$ is the energy). 
Although almost all of the bands are entangled with each other, some bands are disentangled from the others. 
In order to focus on the topological property of the 2$Q$ MC state, we examine the 1/8- and 3/8-filling cases, where the 16th and 48th bands from the lowest energy level are drawn in red and blue in Fig.~\ref{fig: band}(c), respectively. 

To investigate the topological property, we calculate the Chern number $C_n$ for the $n$th band by using the formula in Ref.~\cite{fukui2005chern}. 
We show the Chern number for the $n$th band up to the 64th band in Fig.~\ref{fig: band}(d). 
The behavior of $C_n$ is complicated, but one can see nonzero integer Chern numbers depending on the band, for example, $C_{16}=-1$ and $C_{48}=-1$ corresponding to the 1/8 and 3/8 fillings, respectively. 
From this result, a different feature from the SkX is found. 
In the SkX, the nonzero Chern numbers appear from the lowest-energy band~\cite{Hamamoto_PhysRevB.92.115417, Gobel_PhysRevB.95.094413, Gobel_PhysRevB.96.060406}, while the Chern number becomes zero for the lowest-energy band ($C_0 = 0$) in the 2$Q$ MC state. 
This difference might arise from the difference in their real-space spin textures, where the continuum limit can be taken for the spin texture of the SkX, while it cannot be for that of the 2$Q$ MC state. 
In this context, the 2$Q$ MC state might also exhibit different behavior of the topological Hall effect in the metallic case~\cite{tatara2002chirality, nakazawa2014effects, ishizuka2018spin, nakazawa2018topological, Denisov_PhysRevB.98.195439, Rosales_PhysRevB.99.035163, Mohanta_PhysRevB.100.064429, Bouaziz_PhysRevLett.126.147203, Matsui_PhysRevB.104.174432}. 
Such an analysis will be left as an intriguing issue for future study. 

The summation of the Chern number up to the Fermi level, which is also shown in Fig.~\ref{fig: band}(d), corresponds to the quantized Hall conductivity in the unit of $e^2/h$. 
The Hall conductance is calculated by using the Kubo formula as
\begin{align}
\label{eq: sigma}
\sigma_{xy}=&- \frac{i e^2}{h}\frac{2\pi}{N} \sum_{n \bm{k}} f(\varepsilon_{n\bm{k}}) \cr
& \times \sum_{m (\neq n)} \frac{ \bra{n \bm{k}} \tilde{J}_x \ket{m \bm{k}}\bra{m \bm{k}}\tilde{J}_y\ket{n \bm{k}} - (n \leftrightarrow m)}{(\varepsilon_{n\bm{k}}-\varepsilon_{m\bm{k}})^2}, \cr
\end{align}
where $n, m$ are the band indices, $\varepsilon_{n\bm{k}}$ and $\ket{n \bm{k}}$ are the eigenvalues and eigenstates of the Hamiltonian in Eq.~(\ref{eq: HamKLM}), $f(\varepsilon_{n\bm{k}})$ is the Fermi distribution function, and $\tilde{J}_\eta= \partial H/\partial k_\eta$ for $\eta=x,y$ is the current operator. 
When the system is insulating and the Fermi energy is located between the $n$th and $n+1$th bands, $\sigma_{xy}$ at zero temperature is also represented by the Chern number as 
\begin{align}
\sigma_{xy}= \frac{e^2}{h}C^{\rm total}_n. 
\end{align}
Thus, $\sigma_{xy}=2e^2/h$ for the 1/8 filling and $\sigma_{xy}=-2e^2/h$ for the 3/8 filling; we indeed obtain these values by directly performing the numerical evaluation of Eq.~(\ref{eq: sigma}).

Moreover, the 2$Q$ MC state exhibits the antisymmetric spin splitting in the band structure. 
We show the $x$ and $y$ components of the momentum-resolved spin polarization, $s^{x,y}_{\bm{k}}$, in Figs.~\ref{fig: band_spinsplit}(a) and \ref{fig: band_spinsplit}(b), respectively. 
As shown in Fig.~\ref{fig: band_spinsplit}, the antisymmetric spin polarization mainly appears in the ${\rm X}_1$--$\Gamma$--${\rm X}_2$ line for the $x$-spin component and in the ${\rm M}_1$--$\Gamma$--${\rm M}_2$ line for both $x$- and $y$-spin components. 
Since the functional form is approximately expressed as $\bm{k} \times \bm{s}_{\bm{k}}$ in the limit of $\bm{k}\to \bm{0}$, this antisymmetric spin splitting is categorized into the polar type like the Rashba metal~\cite{rashba1960properties, comment_antisymmetric_2QMC}. 
This indicates that the spin structure of the 2$Q$ MC state in Fig.~\ref{fig: band}(a) is approximately represented by the double-$Q$ superposition of out-of-plane cycloidal spiral structure, whose spiral plane lies on the parallel to $\bm{Q}_\nu$ and $z$ direction~\cite{Hayami_PhysRevB.105.024413}. 
Indeed, such a tendency of the spiral plane was confirmed by looking into the spin structure factor (not shown). 
It is noted that this antisymmetric spin splitting is caused by the noncollinear spin textures rather than the relativistic spin--orbit coupling in contrast to the Rashba metal, since the model Hamiltonian just includes the spin-independent hopping term~\cite{Hayami_PhysRevB.101.220403, Hayami_PhysRevB.102.144441}. 
In addition, the 2$Q$ MC state exhibits the ferroaxial nature, since the spiral plane is usually represented by the linear combination of the out-of-plane cycloidal and proper-screw spirals owing to their degeneracy in the present model, which results in breaking the mirror symmetry along the $z$ direction~\cite{Hayami_PhysRevB.106.144402}.

\subsection{Edge state}
\label{sec: Edge state}

\begin{figure}[tb!]
\begin{center}
\includegraphics[width=1.0\hsize]{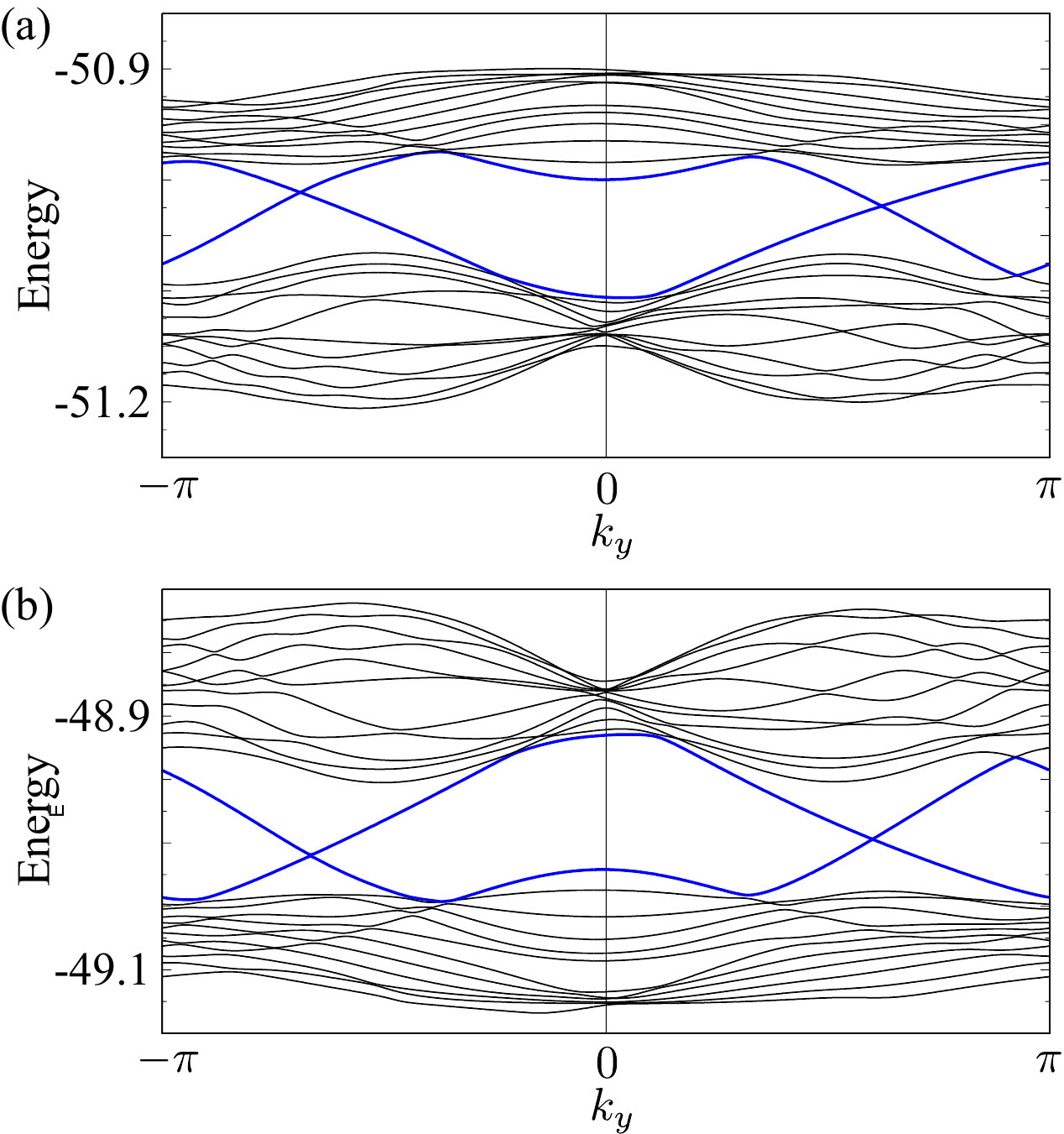} 
\caption{
\label{fig: surface} 
Energy dispersions with the (100) edges close to the region of (a) the 1/8 and (b) 3/8 fillings. 
The model parameters are the same as those in Fig.~\ref{fig: band}(c). 
The thick blue lines represent chiral edge states traversing the gaps, which arise from the nonzero Chern number in the 2$Q$ MC state. 
}
\end{center}
\end{figure}

Finally, let us present the edge state around the 1/8- and 3/8-filling region, where the band gap opens, as shown in Fig.~\ref{fig: band}(c). 
Figures~\ref{fig: surface}(a) and \ref{fig: surface}(b) show the energy dispersion with the (100) edges close to the 1/8 and 3/8 fillings, respectively; the band localized with the (100) edges are drawn in blue.  
Both dispersions exhibit the chiral edge states traversing the gaps, whose number equals the summation of the Chern number, i.e., the bulk-edge correspondence. 
In the end, the 2$Q$ MC state corresponds to the Chern insulating state at 1/8 and 3/8 fillings like the square SkX. 

\section{Summary}
\label{sec: Summary}

To summarize, we have investigated the topologically nontrivial multiple-$Q$ state with a particular focus on the symmetry of the constituent ordering wave vectors in momentum space. 
We found that the unconventional double-$Q$ magnetic chiral (2$Q$ MC) state appears when the ordering wave vectors lie on the Brillouin zone boundary. 
By performing the simulated annealing for the spin model in momentum space, we have shown that the 2$Q$ MC state is stabilized by the interplay between the easy-axis anisotropic interaction and the high-harmonic wave-vector interaction under the external magnetic field. 
The obtained noncoplanar spin texture in the 2$Q$ MC state exhibits nonzero uniform scalar chirality, while it is not connected to that in the SkX owing to the different symmetry lines in the Brillouin zone for the ordering wave vectors. 
In spite of the difference in the spin texture from the SkX, the 2$Q$ MC state becomes the Chern insulating state when the coupling to an itinerant electron is considered. 
We have shown the bulk-edge correspondence in the 2$Q$ MC state. 
Our results provide a possibility of further exotic multiple-$Q$ states with a nontrivial topological spin texture by considering the degree of freedom in terms of the position of the constituent ordering wave vectors in momentum space.

\begin{acknowledgments}
This research was supported by JSPS KAKENHI Grants Numbers JP21H01037, JP22H04468, JP22H00101, JP22H01183, JP23K03288, JP23H04869, and by JST PRESTO (JPMJPR20L8). 
Parts of the numerical calculations were performed in the supercomputing systems in ISSP, the University of Tokyo.
\end{acknowledgments}

\appendix
\section{Phase diagram without high-harmonic wave-vector interactions}
\label{appendix}

\begin{figure}[tb!]
\begin{center}
\includegraphics[width=1.0\hsize]{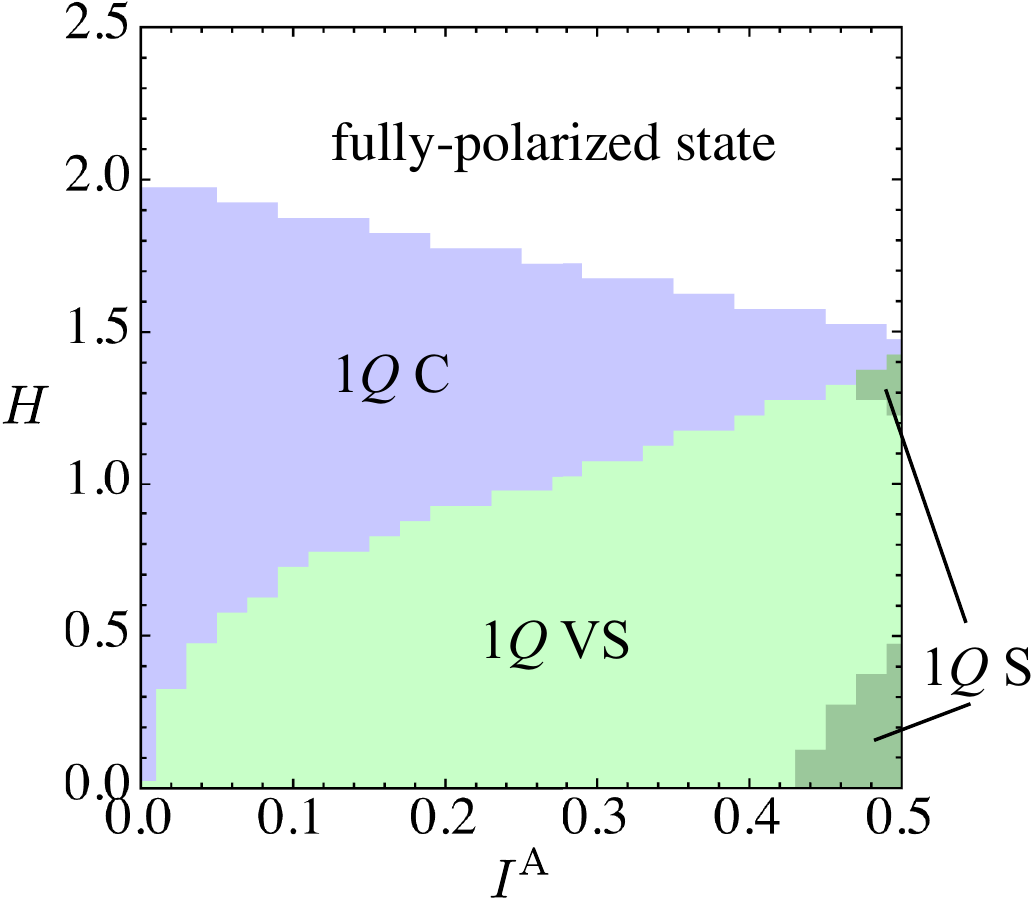} 
\caption{
\label{fig: PD_app} 
Ground-state phase diagram of the model in Eq.~(\ref{eq: Ham}) on the square lattice at $J'=0$. 
}
\end{center}
\end{figure}

In this Appendix, we show the phase diagram of the model in Eq.~(\ref{eq: Ham}) in the absence of the high-harmonic wave-vector interaction, i.e., $J'=0$. 
Figure~\ref{fig: PD_app} shows the ground-state phase diagram as functions of $I^{\rm A}$ and $H$. 
Comparing the phase diagram with $J'=0.6$ in Fig.~\ref{fig: PD}, no double-$Q$ states appear in the phase diagram; $1Q$ VS state is stabilized for small $H$, 1$Q$ C state is stabilized for large $H$ and small $I^{\rm A}$, and the 1$Q$ S state is stabilized for small $H$ and large $I^{\rm A}$. 
Thus, the high-harmonic wave-vector interaction plays an important role in inducing the double-$Q$ states including the 2$Q$ MC state.

\bibliographystyle{apsrev}
\bibliography{../ref}

\begin{thebibliography}{114}
\expandafter\ifx\csname natexlab\endcsname\relax\def\natexlab#1{#1}\fi
\expandafter\ifx\csname bibnamefont\endcsname\relax
  \def\bibnamefont#1{#1}\fi
\expandafter\ifx\csname bibfnamefont\endcsname\relax
  \def\bibfnamefont#1{#1}\fi
\expandafter\ifx\csname citenamefont\endcsname\relax
  \def\citenamefont#1{#1}\fi
\expandafter\ifx\csname url\endcsname\relax
  \def\url#1{\texttt{#1}}\fi
\expandafter\ifx\csname urlprefix\endcsname\relax\def\urlprefix{URL }\fi
\providecommand{\bibinfo}[2]{#2}
\providecommand{\eprint}[2][]{\url{#2}}

\bibitem[{\citenamefont{Haldane}(1988)}]{Haldane_PhysRevLett.61.2015}
\bibinfo{author}{\bibfnamefont{F.~D.~M.} \bibnamefont{Haldane}},
  \bibinfo{journal}{Phys. Rev. Lett.} \textbf{\bibinfo{volume}{61}},
  \bibinfo{pages}{2015} (\bibinfo{year}{1988}).

\bibitem[{\citenamefont{Thouless et~al.}(1982)\citenamefont{Thouless, Kohmoto,
  Nightingale, and den Nijs}}]{Thouless_PhysRevLett.49.405}
\bibinfo{author}{\bibfnamefont{D.~J.} \bibnamefont{Thouless}},
  \bibinfo{author}{\bibfnamefont{M.}~\bibnamefont{Kohmoto}},
  \bibinfo{author}{\bibfnamefont{M.~P.} \bibnamefont{Nightingale}},
  \bibnamefont{and} \bibinfo{author}{\bibfnamefont{M.}~\bibnamefont{den Nijs}},
  \bibinfo{journal}{Phys. Rev. Lett.} \textbf{\bibinfo{volume}{49}},
  \bibinfo{pages}{405} (\bibinfo{year}{1982}).

\bibitem[{\citenamefont{Kohmoto}(1985)}]{kohmoto1985topological}
\bibinfo{author}{\bibfnamefont{M.}~\bibnamefont{Kohmoto}},
  \bibinfo{journal}{Ann. Phys.} \textbf{\bibinfo{volume}{160}},
  \bibinfo{pages}{343} (\bibinfo{year}{1985}).

\bibitem[{\citenamefont{Berry}(1984)}]{berry1984quantal}
\bibinfo{author}{\bibfnamefont{M.~V.} \bibnamefont{Berry}},
  \bibinfo{journal}{Proceedings of the Royal Society of London A: Mathematical,
  Physical and Engineering Sciences} \textbf{\bibinfo{volume}{392}},
  \bibinfo{pages}{45} (\bibinfo{year}{1984}).

\bibitem[{\citenamefont{Loss and Goldbart}(1992)}]{Loss_PhysRevB.45.13544}
\bibinfo{author}{\bibfnamefont{D.}~\bibnamefont{Loss}} \bibnamefont{and}
  \bibinfo{author}{\bibfnamefont{P.~M.} \bibnamefont{Goldbart}},
  \bibinfo{journal}{Phys. Rev. B} \textbf{\bibinfo{volume}{45}},
  \bibinfo{pages}{13544} (\bibinfo{year}{1992}).

\bibitem[{\citenamefont{Ye et~al.}(1999)\citenamefont{Ye, Kim, Millis,
  Shraiman, Majumdar, and Te\ifmmode \check{s}\else
  \v{s}\fi{}anovi\ifmmode~\acute{c}\else \'{c}\fi{}}}]{Ye_PhysRevLett.83.3737}
\bibinfo{author}{\bibfnamefont{J.}~\bibnamefont{Ye}},
  \bibinfo{author}{\bibfnamefont{Y.~B.} \bibnamefont{Kim}},
  \bibinfo{author}{\bibfnamefont{A.~J.} \bibnamefont{Millis}},
  \bibinfo{author}{\bibfnamefont{B.~I.} \bibnamefont{Shraiman}},
  \bibinfo{author}{\bibfnamefont{P.}~\bibnamefont{Majumdar}}, \bibnamefont{and}
  \bibinfo{author}{\bibfnamefont{Z.}~\bibnamefont{Te\ifmmode \check{s}\else
  \v{s}\fi{}anovi\ifmmode~\acute{c}\else \'{c}\fi{}}}, \bibinfo{journal}{Phys.
  Rev. Lett.} \textbf{\bibinfo{volume}{83}}, \bibinfo{pages}{3737}
  (\bibinfo{year}{1999}).

\bibitem[{\citenamefont{Ohgushi et~al.}(2000)\citenamefont{Ohgushi, Murakami,
  and Nagaosa}}]{Ohgushi_PhysRevB.62.R6065}
\bibinfo{author}{\bibfnamefont{K.}~\bibnamefont{Ohgushi}},
  \bibinfo{author}{\bibfnamefont{S.}~\bibnamefont{Murakami}}, \bibnamefont{and}
  \bibinfo{author}{\bibfnamefont{N.}~\bibnamefont{Nagaosa}},
  \bibinfo{journal}{Phys. Rev. B} \textbf{\bibinfo{volume}{62}},
  \bibinfo{pages}{R6065} (\bibinfo{year}{2000}).

\bibitem[{\citenamefont{Shindou and
  Nagaosa}(2001)}]{Shindou_PhysRevLett.87.116801}
\bibinfo{author}{\bibfnamefont{R.}~\bibnamefont{Shindou}} \bibnamefont{and}
  \bibinfo{author}{\bibfnamefont{N.}~\bibnamefont{Nagaosa}},
  \bibinfo{journal}{Phys. Rev. Lett.} \textbf{\bibinfo{volume}{87}},
  \bibinfo{pages}{116801} (\bibinfo{year}{2001}).

\bibitem[{\citenamefont{Nagaosa et~al.}(2010)\citenamefont{Nagaosa, Sinova,
  Onoda, MacDonald, and Ong}}]{Nagaosa_RevModPhys.82.1539}
\bibinfo{author}{\bibfnamefont{N.}~\bibnamefont{Nagaosa}},
  \bibinfo{author}{\bibfnamefont{J.}~\bibnamefont{Sinova}},
  \bibinfo{author}{\bibfnamefont{S.}~\bibnamefont{Onoda}},
  \bibinfo{author}{\bibfnamefont{A.~H.} \bibnamefont{MacDonald}},
  \bibnamefont{and} \bibinfo{author}{\bibfnamefont{N.~P.} \bibnamefont{Ong}},
  \bibinfo{journal}{Rev. Mod. Phys.} \textbf{\bibinfo{volume}{82}},
  \bibinfo{pages}{1539} (\bibinfo{year}{2010}).

\bibitem[{\citenamefont{Hayami and
  Motome}(2021{\natexlab{a}})}]{hayami2021topological}
\bibinfo{author}{\bibfnamefont{S.}~\bibnamefont{Hayami}} \bibnamefont{and}
  \bibinfo{author}{\bibfnamefont{Y.}~\bibnamefont{Motome}},
  \bibinfo{journal}{J. Phys.: Condens. Matter} \textbf{\bibinfo{volume}{33}},
  \bibinfo{pages}{443001} (\bibinfo{year}{2021}{\natexlab{a}}).

\bibitem[{\citenamefont{Momoi et~al.}(1997)\citenamefont{Momoi, Kubo, and
  Niki}}]{Momoi_PhysRevLett.79.2081}
\bibinfo{author}{\bibfnamefont{T.}~\bibnamefont{Momoi}},
  \bibinfo{author}{\bibfnamefont{K.}~\bibnamefont{Kubo}}, \bibnamefont{and}
  \bibinfo{author}{\bibfnamefont{K.}~\bibnamefont{Niki}},
  \bibinfo{journal}{Phys. Rev. Lett.} \textbf{\bibinfo{volume}{79}},
  \bibinfo{pages}{2081} (\bibinfo{year}{1997}).

\bibitem[{\citenamefont{Martin and
  Batista}(2008)}]{Martin_PhysRevLett.101.156402}
\bibinfo{author}{\bibfnamefont{I.}~\bibnamefont{Martin}} \bibnamefont{and}
  \bibinfo{author}{\bibfnamefont{C.~D.} \bibnamefont{Batista}},
  \bibinfo{journal}{Phys. Rev. Lett.} \textbf{\bibinfo{volume}{101}},
  \bibinfo{pages}{156402} (\bibinfo{year}{2008}).

\bibitem[{\citenamefont{Chern and
  Batista}(2012)}]{Chern_PhysRevLett.109.156801}
\bibinfo{author}{\bibfnamefont{G.-W.} \bibnamefont{Chern}} \bibnamefont{and}
  \bibinfo{author}{\bibfnamefont{C.~D.} \bibnamefont{Batista}},
  \bibinfo{journal}{Phys. Rev. Lett.} \textbf{\bibinfo{volume}{109}},
  \bibinfo{pages}{156801} (\bibinfo{year}{2012}).

\bibitem[{\citenamefont{Venderbos
  et~al.}(2012{\natexlab{a}})\citenamefont{Venderbos, Kourtis, van~den Brink,
  and Daghofer}}]{Venderbos_PhysRevLett.108.126405}
\bibinfo{author}{\bibfnamefont{J.~W.~F.} \bibnamefont{Venderbos}},
  \bibinfo{author}{\bibfnamefont{S.}~\bibnamefont{Kourtis}},
  \bibinfo{author}{\bibfnamefont{J.}~\bibnamefont{van~den Brink}},
  \bibnamefont{and} \bibinfo{author}{\bibfnamefont{M.}~\bibnamefont{Daghofer}},
  \bibinfo{journal}{Phys. Rev. Lett.} \textbf{\bibinfo{volume}{108}},
  \bibinfo{pages}{126405} (\bibinfo{year}{2012}{\natexlab{a}}).

\bibitem[{\citenamefont{Akagi and Motome}(2010)}]{Akagi_JPSJ.79.083711}
\bibinfo{author}{\bibfnamefont{Y.}~\bibnamefont{Akagi}} \bibnamefont{and}
  \bibinfo{author}{\bibfnamefont{Y.}~\bibnamefont{Motome}},
  \bibinfo{journal}{J. Phys. Soc. Jpn.} \textbf{\bibinfo{volume}{79}},
  \bibinfo{pages}{083711} (\bibinfo{year}{2010}).

\bibitem[{\citenamefont{Akagi et~al.}(2012)\citenamefont{Akagi, Udagawa, and
  Motome}}]{Akagi_PhysRevLett.108.096401}
\bibinfo{author}{\bibfnamefont{Y.}~\bibnamefont{Akagi}},
  \bibinfo{author}{\bibfnamefont{M.}~\bibnamefont{Udagawa}}, \bibnamefont{and}
  \bibinfo{author}{\bibfnamefont{Y.}~\bibnamefont{Motome}},
  \bibinfo{journal}{Phys. Rev. Lett.} \textbf{\bibinfo{volume}{108}},
  \bibinfo{pages}{096401} (\bibinfo{year}{2012}).

\bibitem[{\citenamefont{Hayami and Motome}(2014)}]{Hayami_PhysRevB.90.060402}
\bibinfo{author}{\bibfnamefont{S.}~\bibnamefont{Hayami}} \bibnamefont{and}
  \bibinfo{author}{\bibfnamefont{Y.}~\bibnamefont{Motome}},
  \bibinfo{journal}{Phys. Rev. B} \textbf{\bibinfo{volume}{90}},
  \bibinfo{pages}{060402(R)} (\bibinfo{year}{2014}).

\bibitem[{\citenamefont{Hayami and Motome}(2015)}]{hayami_PhysRevB.91.075104}
\bibinfo{author}{\bibfnamefont{S.}~\bibnamefont{Hayami}} \bibnamefont{and}
  \bibinfo{author}{\bibfnamefont{Y.}~\bibnamefont{Motome}},
  \bibinfo{journal}{Phys. Rev. B} \textbf{\bibinfo{volume}{91}},
  \bibinfo{pages}{075104} (\bibinfo{year}{2015}).

\bibitem[{\citenamefont{Hayami et~al.}(2016{\natexlab{a}})\citenamefont{Hayami,
  Ozawa, and Motome}}]{Hayami_PhysRevB.94.024424}
\bibinfo{author}{\bibfnamefont{S.}~\bibnamefont{Hayami}},
  \bibinfo{author}{\bibfnamefont{R.}~\bibnamefont{Ozawa}}, \bibnamefont{and}
  \bibinfo{author}{\bibfnamefont{Y.}~\bibnamefont{Motome}},
  \bibinfo{journal}{Phys. Rev. B} \textbf{\bibinfo{volume}{94}},
  \bibinfo{pages}{024424} (\bibinfo{year}{2016}{\natexlab{a}}).

\bibitem[{\citenamefont{Huang et~al.}(2020)\citenamefont{Huang, Dong, Kotetes,
  and Zhou}}]{Huang_PhysRevB.102.195120}
\bibinfo{author}{\bibfnamefont{Y.-P.} \bibnamefont{Huang}},
  \bibinfo{author}{\bibfnamefont{J.-W.} \bibnamefont{Dong}},
  \bibinfo{author}{\bibfnamefont{P.}~\bibnamefont{Kotetes}}, \bibnamefont{and}
  \bibinfo{author}{\bibfnamefont{S.}~\bibnamefont{Zhou}},
  \bibinfo{journal}{Phys. Rev. B} \textbf{\bibinfo{volume}{102}},
  \bibinfo{pages}{195120} (\bibinfo{year}{2020}).

\bibitem[{\citenamefont{Venderbos
  et~al.}(2012{\natexlab{b}})\citenamefont{Venderbos, Daghofer, van~den Brink,
  and Kumar}}]{Venderbos_PhysRevLett.109.166405}
\bibinfo{author}{\bibfnamefont{J.~W.~F.} \bibnamefont{Venderbos}},
  \bibinfo{author}{\bibfnamefont{M.}~\bibnamefont{Daghofer}},
  \bibinfo{author}{\bibfnamefont{J.}~\bibnamefont{van~den Brink}},
  \bibnamefont{and} \bibinfo{author}{\bibfnamefont{S.}~\bibnamefont{Kumar}},
  \bibinfo{journal}{Phys. Rev. Lett.} \textbf{\bibinfo{volume}{109}},
  \bibinfo{pages}{166405} (\bibinfo{year}{2012}{\natexlab{b}}).

\bibitem[{\citenamefont{Jiang et~al.}(2015)\citenamefont{Jiang, Zhang, Zhou,
  and Wang}}]{Jiang_PhysRevLett.114.216402}
\bibinfo{author}{\bibfnamefont{K.}~\bibnamefont{Jiang}},
  \bibinfo{author}{\bibfnamefont{Y.}~\bibnamefont{Zhang}},
  \bibinfo{author}{\bibfnamefont{S.}~\bibnamefont{Zhou}}, \bibnamefont{and}
  \bibinfo{author}{\bibfnamefont{Z.}~\bibnamefont{Wang}},
  \bibinfo{journal}{Phys. Rev. Lett.} \textbf{\bibinfo{volume}{114}},
  \bibinfo{pages}{216402} (\bibinfo{year}{2015}).

\bibitem[{\citenamefont{Barros et~al.}(2014)\citenamefont{Barros, Venderbos,
  Chern, and Batista}}]{Barros_PhysRevB.90.245119}
\bibinfo{author}{\bibfnamefont{K.}~\bibnamefont{Barros}},
  \bibinfo{author}{\bibfnamefont{J.~W.~F.} \bibnamefont{Venderbos}},
  \bibinfo{author}{\bibfnamefont{G.-W.} \bibnamefont{Chern}}, \bibnamefont{and}
  \bibinfo{author}{\bibfnamefont{C.~D.} \bibnamefont{Batista}},
  \bibinfo{journal}{Phys. Rev. B} \textbf{\bibinfo{volume}{90}},
  \bibinfo{pages}{245119} (\bibinfo{year}{2014}).

\bibitem[{\citenamefont{Ghosh et~al.}(2016)\citenamefont{Ghosh, O'Brien,
  Henley, and Lawler}}]{Ghosh_PhysRevB.93.024401}
\bibinfo{author}{\bibfnamefont{S.}~\bibnamefont{Ghosh}},
  \bibinfo{author}{\bibfnamefont{P.}~\bibnamefont{O'Brien}},
  \bibinfo{author}{\bibfnamefont{C.~L.} \bibnamefont{Henley}},
  \bibnamefont{and} \bibinfo{author}{\bibfnamefont{M.~J.}
  \bibnamefont{Lawler}}, \bibinfo{journal}{Phys. Rev. B}
  \textbf{\bibinfo{volume}{93}}, \bibinfo{pages}{024401}
  (\bibinfo{year}{2016}).

\bibitem[{\citenamefont{Nagaosa and Tokura}(2013)}]{nagaosa2013topological}
\bibinfo{author}{\bibfnamefont{N.}~\bibnamefont{Nagaosa}} \bibnamefont{and}
  \bibinfo{author}{\bibfnamefont{Y.}~\bibnamefont{Tokura}},
  \bibinfo{journal}{Nat. Nanotechnol.} \textbf{\bibinfo{volume}{8}},
  \bibinfo{pages}{899} (\bibinfo{year}{2013}).

\bibitem[{\citenamefont{Dzyaloshinsky}(1958)}]{dzyaloshinsky1958thermodynamic}
\bibinfo{author}{\bibfnamefont{I.}~\bibnamefont{Dzyaloshinsky}},
  \bibinfo{journal}{J. Phys. Chem. Solids} \textbf{\bibinfo{volume}{4}},
  \bibinfo{pages}{241} (\bibinfo{year}{1958}).

\bibitem[{\citenamefont{Moriya}(1960)}]{moriya1960anisotropic}
\bibinfo{author}{\bibfnamefont{T.}~\bibnamefont{Moriya}},
  \bibinfo{journal}{Phys. Rev.} \textbf{\bibinfo{volume}{120}},
  \bibinfo{pages}{91} (\bibinfo{year}{1960}).

\bibitem[{\citenamefont{R{\"o}{\ss}ler
  et~al.}(2006)\citenamefont{R{\"o}{\ss}ler, Bogdanov, and
  Pfleiderer}}]{rossler2006spontaneous}
\bibinfo{author}{\bibfnamefont{U.~K.} \bibnamefont{R{\"o}{\ss}ler}},
  \bibinfo{author}{\bibfnamefont{A.~N.} \bibnamefont{Bogdanov}},
  \bibnamefont{and}
  \bibinfo{author}{\bibfnamefont{C.}~\bibnamefont{Pfleiderer}},
  \bibinfo{journal}{Nature} \textbf{\bibinfo{volume}{442}},
  \bibinfo{pages}{797} (\bibinfo{year}{2006}).

\bibitem[{\citenamefont{Yi et~al.}(2009)\citenamefont{Yi, Onoda, Nagaosa, and
  Han}}]{Yi_PhysRevB.80.054416}
\bibinfo{author}{\bibfnamefont{S.~D.} \bibnamefont{Yi}},
  \bibinfo{author}{\bibfnamefont{S.}~\bibnamefont{Onoda}},
  \bibinfo{author}{\bibfnamefont{N.}~\bibnamefont{Nagaosa}}, \bibnamefont{and}
  \bibinfo{author}{\bibfnamefont{J.~H.} \bibnamefont{Han}},
  \bibinfo{journal}{Phys. Rev. B} \textbf{\bibinfo{volume}{80}},
  \bibinfo{pages}{054416} (\bibinfo{year}{2009}).

\bibitem[{\citenamefont{Butenko et~al.}(2010)\citenamefont{Butenko, Leonov,
  R\"o\ss{}ler, and Bogdanov}}]{Butenko_PhysRevB.82.052403}
\bibinfo{author}{\bibfnamefont{A.~B.} \bibnamefont{Butenko}},
  \bibinfo{author}{\bibfnamefont{A.~A.} \bibnamefont{Leonov}},
  \bibinfo{author}{\bibfnamefont{U.~K.} \bibnamefont{R\"o\ss{}ler}},
  \bibnamefont{and} \bibinfo{author}{\bibfnamefont{A.~N.}
  \bibnamefont{Bogdanov}}, \bibinfo{journal}{Phys. Rev. B}
  \textbf{\bibinfo{volume}{82}}, \bibinfo{pages}{052403}
  (\bibinfo{year}{2010}).

\bibitem[{\citenamefont{Hayami}(2022{\natexlab{a}})}]{Hayami_PhysRevB.105.014408}
\bibinfo{author}{\bibfnamefont{S.}~\bibnamefont{Hayami}},
  \bibinfo{journal}{Phys. Rev. B} \textbf{\bibinfo{volume}{105}},
  \bibinfo{pages}{014408} (\bibinfo{year}{2022}{\natexlab{a}}).

\bibitem[{\citenamefont{Lin}(2021)}]{lin2021skyrmion}
\bibinfo{author}{\bibfnamefont{S.-Z.} \bibnamefont{Lin}},
  \bibinfo{journal}{arXiv:2112.12850}  (\bibinfo{year}{2021}).

\bibitem[{\citenamefont{Okubo et~al.}(2012)\citenamefont{Okubo, Chung, and
  Kawamura}}]{Okubo_PhysRevLett.108.017206}
\bibinfo{author}{\bibfnamefont{T.}~\bibnamefont{Okubo}},
  \bibinfo{author}{\bibfnamefont{S.}~\bibnamefont{Chung}}, \bibnamefont{and}
  \bibinfo{author}{\bibfnamefont{H.}~\bibnamefont{Kawamura}},
  \bibinfo{journal}{Phys. Rev. Lett.} \textbf{\bibinfo{volume}{108}},
  \bibinfo{pages}{017206} (\bibinfo{year}{2012}).

\bibitem[{\citenamefont{Leonov and Mostovoy}(2015)}]{leonov2015multiply}
\bibinfo{author}{\bibfnamefont{A.~O.} \bibnamefont{Leonov}} \bibnamefont{and}
  \bibinfo{author}{\bibfnamefont{M.}~\bibnamefont{Mostovoy}},
  \bibinfo{journal}{Nat. Commun.} \textbf{\bibinfo{volume}{6}},
  \bibinfo{pages}{8275} (\bibinfo{year}{2015}).

\bibitem[{\citenamefont{Lin and Hayami}(2016)}]{Lin_PhysRevB.93.064430}
\bibinfo{author}{\bibfnamefont{S.-Z.} \bibnamefont{Lin}} \bibnamefont{and}
  \bibinfo{author}{\bibfnamefont{S.}~\bibnamefont{Hayami}},
  \bibinfo{journal}{Phys. Rev. B} \textbf{\bibinfo{volume}{93}},
  \bibinfo{pages}{064430} (\bibinfo{year}{2016}).

\bibitem[{\citenamefont{Hayami et~al.}(2016{\natexlab{b}})\citenamefont{Hayami,
  Lin, and Batista}}]{Hayami_PhysRevB.93.184413}
\bibinfo{author}{\bibfnamefont{S.}~\bibnamefont{Hayami}},
  \bibinfo{author}{\bibfnamefont{S.-Z.} \bibnamefont{Lin}}, \bibnamefont{and}
  \bibinfo{author}{\bibfnamefont{C.~D.} \bibnamefont{Batista}},
  \bibinfo{journal}{Phys. Rev. B} \textbf{\bibinfo{volume}{93}},
  \bibinfo{pages}{184413} (\bibinfo{year}{2016}{\natexlab{b}}).

\bibitem[{\citenamefont{Hayami et~al.}(2016{\natexlab{c}})\citenamefont{Hayami,
  Lin, Kamiya, and Batista}}]{Hayami_PhysRevB.94.174420}
\bibinfo{author}{\bibfnamefont{S.}~\bibnamefont{Hayami}},
  \bibinfo{author}{\bibfnamefont{S.-Z.} \bibnamefont{Lin}},
  \bibinfo{author}{\bibfnamefont{Y.}~\bibnamefont{Kamiya}}, \bibnamefont{and}
  \bibinfo{author}{\bibfnamefont{C.~D.} \bibnamefont{Batista}},
  \bibinfo{journal}{Phys. Rev. B} \textbf{\bibinfo{volume}{94}},
  \bibinfo{pages}{174420} (\bibinfo{year}{2016}{\natexlab{c}}).

\bibitem[{\citenamefont{Lin and Batista}(2018)}]{Lin_PhysRevLett.120.077202}
\bibinfo{author}{\bibfnamefont{S.-Z.} \bibnamefont{Lin}} \bibnamefont{and}
  \bibinfo{author}{\bibfnamefont{C.~D.} \bibnamefont{Batista}},
  \bibinfo{journal}{Phys. Rev. Lett.} \textbf{\bibinfo{volume}{120}},
  \bibinfo{pages}{077202} (\bibinfo{year}{2018}).

\bibitem[{\citenamefont{Hayami}(2021)}]{Hayami_PhysRevB.103.224418}
\bibinfo{author}{\bibfnamefont{S.}~\bibnamefont{Hayami}},
  \bibinfo{journal}{Phys. Rev. B} \textbf{\bibinfo{volume}{103}},
  \bibinfo{pages}{224418} (\bibinfo{year}{2021}).

\bibitem[{\citenamefont{Utesov}(2021)}]{Utesov_PhysRevB.103.064414}
\bibinfo{author}{\bibfnamefont{O.~I.} \bibnamefont{Utesov}},
  \bibinfo{journal}{Phys. Rev. B} \textbf{\bibinfo{volume}{103}},
  \bibinfo{pages}{064414} (\bibinfo{year}{2021}).

\bibitem[{\citenamefont{Utesov}(2022)}]{Utesov_PhysRevB.105.054435}
\bibinfo{author}{\bibfnamefont{O.~I.} \bibnamefont{Utesov}},
  \bibinfo{journal}{Phys. Rev. B} \textbf{\bibinfo{volume}{105}},
  \bibinfo{pages}{054435} (\bibinfo{year}{2022}).

\bibitem[{\citenamefont{Amoroso et~al.}(2020)\citenamefont{Amoroso, Barone, and
  Picozzi}}]{amoroso2020spontaneous}
\bibinfo{author}{\bibfnamefont{D.}~\bibnamefont{Amoroso}},
  \bibinfo{author}{\bibfnamefont{P.}~\bibnamefont{Barone}}, \bibnamefont{and}
  \bibinfo{author}{\bibfnamefont{S.}~\bibnamefont{Picozzi}},
  \bibinfo{journal}{Nat. Commun.} \textbf{\bibinfo{volume}{11}},
  \bibinfo{pages}{5784} (\bibinfo{year}{2020}).

\bibitem[{\citenamefont{Yambe and Hayami}(2021)}]{yambe2021skyrmion}
\bibinfo{author}{\bibfnamefont{R.}~\bibnamefont{Yambe}} \bibnamefont{and}
  \bibinfo{author}{\bibfnamefont{S.}~\bibnamefont{Hayami}},
  \bibinfo{journal}{Sci. Rep.} \textbf{\bibinfo{volume}{11}},
  \bibinfo{pages}{11184} (\bibinfo{year}{2021}).

\bibitem[{\citenamefont{Amoroso et~al.}(2021)\citenamefont{Amoroso, Barone, and
  Picozzi}}]{amoroso2021tuning}
\bibinfo{author}{\bibfnamefont{D.}~\bibnamefont{Amoroso}},
  \bibinfo{author}{\bibfnamefont{P.}~\bibnamefont{Barone}}, \bibnamefont{and}
  \bibinfo{author}{\bibfnamefont{S.}~\bibnamefont{Picozzi}},
  \bibinfo{journal}{Nanomaterials} \textbf{\bibinfo{volume}{11}},
  \bibinfo{pages}{1873} (\bibinfo{year}{2021}).

\bibitem[{\citenamefont{Hirschberger et~al.}(2021)\citenamefont{Hirschberger,
  Hayami, and Tokura}}]{Hirschberger_10.1088/1367-2630/abdef9}
\bibinfo{author}{\bibfnamefont{M.}~\bibnamefont{Hirschberger}},
  \bibinfo{author}{\bibfnamefont{S.}~\bibnamefont{Hayami}}, \bibnamefont{and}
  \bibinfo{author}{\bibfnamefont{Y.}~\bibnamefont{Tokura}},
  \bibinfo{journal}{New J. Phys.} \textbf{\bibinfo{volume}{23}},
  \bibinfo{pages}{023039} (\bibinfo{year}{2021}).

\bibitem[{\citenamefont{Hayami and
  Motome}(2021{\natexlab{b}})}]{Hayami_PhysRevB.103.054422}
\bibinfo{author}{\bibfnamefont{S.}~\bibnamefont{Hayami}} \bibnamefont{and}
  \bibinfo{author}{\bibfnamefont{Y.}~\bibnamefont{Motome}},
  \bibinfo{journal}{Phys. Rev. B} \textbf{\bibinfo{volume}{103}},
  \bibinfo{pages}{054422} (\bibinfo{year}{2021}{\natexlab{b}}).

\bibitem[{\citenamefont{Yambe and Hayami}(2022)}]{Yambe_PhysRevB.106.174437}
\bibinfo{author}{\bibfnamefont{R.}~\bibnamefont{Yambe}} \bibnamefont{and}
  \bibinfo{author}{\bibfnamefont{S.}~\bibnamefont{Hayami}},
  \bibinfo{journal}{Phys. Rev. B} \textbf{\bibinfo{volume}{106}},
  \bibinfo{pages}{174437} (\bibinfo{year}{2022}).

\bibitem[{\citenamefont{Ruderman and Kittel}(1954)}]{Ruderman}
\bibinfo{author}{\bibfnamefont{M.~A.} \bibnamefont{Ruderman}} \bibnamefont{and}
  \bibinfo{author}{\bibfnamefont{C.}~\bibnamefont{Kittel}},
  \bibinfo{journal}{Phys. Rev.} \textbf{\bibinfo{volume}{96}},
  \bibinfo{pages}{99} (\bibinfo{year}{1954}).

\bibitem[{\citenamefont{Kasuya}(1956)}]{Kasuya}
\bibinfo{author}{\bibfnamefont{T.}~\bibnamefont{Kasuya}},
  \bibinfo{journal}{Prog. Theor. Phys.} \textbf{\bibinfo{volume}{16}},
  \bibinfo{pages}{45} (\bibinfo{year}{1956}).

\bibitem[{\citenamefont{Yosida}(1957)}]{Yosida1957}
\bibinfo{author}{\bibfnamefont{K.}~\bibnamefont{Yosida}},
  \bibinfo{journal}{Phys. Rev.} \textbf{\bibinfo{volume}{106}},
  \bibinfo{pages}{893} (\bibinfo{year}{1957}).

\bibitem[{\citenamefont{Ozawa et~al.}(2017)\citenamefont{Ozawa, Hayami, and
  Motome}}]{Ozawa_PhysRevLett.118.147205}
\bibinfo{author}{\bibfnamefont{R.}~\bibnamefont{Ozawa}},
  \bibinfo{author}{\bibfnamefont{S.}~\bibnamefont{Hayami}}, \bibnamefont{and}
  \bibinfo{author}{\bibfnamefont{Y.}~\bibnamefont{Motome}},
  \bibinfo{journal}{Phys. Rev. Lett.} \textbf{\bibinfo{volume}{118}},
  \bibinfo{pages}{147205} (\bibinfo{year}{2017}).

\bibitem[{\citenamefont{Hayami et~al.}(2017)\citenamefont{Hayami, Ozawa, and
  Motome}}]{Hayami_PhysRevB.95.224424}
\bibinfo{author}{\bibfnamefont{S.}~\bibnamefont{Hayami}},
  \bibinfo{author}{\bibfnamefont{R.}~\bibnamefont{Ozawa}}, \bibnamefont{and}
  \bibinfo{author}{\bibfnamefont{Y.}~\bibnamefont{Motome}},
  \bibinfo{journal}{Phys. Rev. B} \textbf{\bibinfo{volume}{95}},
  \bibinfo{pages}{224424} (\bibinfo{year}{2017}).

\bibitem[{\citenamefont{Wang et~al.}(2020)\citenamefont{Wang, Su, Lin, and
  Batista}}]{Wang_PhysRevLett.124.207201}
\bibinfo{author}{\bibfnamefont{Z.}~\bibnamefont{Wang}},
  \bibinfo{author}{\bibfnamefont{Y.}~\bibnamefont{Su}},
  \bibinfo{author}{\bibfnamefont{S.-Z.} \bibnamefont{Lin}}, \bibnamefont{and}
  \bibinfo{author}{\bibfnamefont{C.~D.} \bibnamefont{Batista}},
  \bibinfo{journal}{Phys. Rev. Lett.} \textbf{\bibinfo{volume}{124}},
  \bibinfo{pages}{207201} (\bibinfo{year}{2020}).

\bibitem[{\citenamefont{Hayami and Motome}(2019)}]{Hayami_PhysRevB.99.094420}
\bibinfo{author}{\bibfnamefont{S.}~\bibnamefont{Hayami}} \bibnamefont{and}
  \bibinfo{author}{\bibfnamefont{Y.}~\bibnamefont{Motome}},
  \bibinfo{journal}{Phys. Rev. B} \textbf{\bibinfo{volume}{99}},
  \bibinfo{pages}{094420} (\bibinfo{year}{2019}).

\bibitem[{\citenamefont{Mitsumoto and
  Kawamura}(2021)}]{Mitsumoto_PhysRevB.104.184432}
\bibinfo{author}{\bibfnamefont{K.}~\bibnamefont{Mitsumoto}} \bibnamefont{and}
  \bibinfo{author}{\bibfnamefont{H.}~\bibnamefont{Kawamura}},
  \bibinfo{journal}{Phys. Rev. B} \textbf{\bibinfo{volume}{104}},
  \bibinfo{pages}{184432} (\bibinfo{year}{2021}).

\bibitem[{\citenamefont{Hayami et~al.}(2021)\citenamefont{Hayami, Okubo, and
  Motome}}]{hayami2021phase}
\bibinfo{author}{\bibfnamefont{S.}~\bibnamefont{Hayami}},
  \bibinfo{author}{\bibfnamefont{T.}~\bibnamefont{Okubo}}, \bibnamefont{and}
  \bibinfo{author}{\bibfnamefont{Y.}~\bibnamefont{Motome}},
  \bibinfo{journal}{Nat. Commun.} \textbf{\bibinfo{volume}{12}},
  \bibinfo{pages}{6927} (\bibinfo{year}{2021}).

\bibitem[{\citenamefont{Mitsumoto and
  Kawamura}(2022)}]{Mitsumoto_PhysRevB.105.094427}
\bibinfo{author}{\bibfnamefont{K.}~\bibnamefont{Mitsumoto}} \bibnamefont{and}
  \bibinfo{author}{\bibfnamefont{H.}~\bibnamefont{Kawamura}},
  \bibinfo{journal}{Phys. Rev. B} \textbf{\bibinfo{volume}{105}},
  \bibinfo{pages}{094427} (\bibinfo{year}{2022}).

\bibitem[{\citenamefont{Kobayashi and
  Hayami}(2022)}]{Kobayashi_PhysRevB.106.L140406}
\bibinfo{author}{\bibfnamefont{K.}~\bibnamefont{Kobayashi}} \bibnamefont{and}
  \bibinfo{author}{\bibfnamefont{S.}~\bibnamefont{Hayami}},
  \bibinfo{journal}{Phys. Rev. B} \textbf{\bibinfo{volume}{106}},
  \bibinfo{pages}{L140406} (\bibinfo{year}{2022}).

\bibitem[{\citenamefont{Bouaziz et~al.}(2022)\citenamefont{Bouaziz,
  Mendive-Tapia, Bl\"ugel, and Staunton}}]{Bouaziz_PhysRevLett.128.157206}
\bibinfo{author}{\bibfnamefont{J.}~\bibnamefont{Bouaziz}},
  \bibinfo{author}{\bibfnamefont{E.}~\bibnamefont{Mendive-Tapia}},
  \bibinfo{author}{\bibfnamefont{S.}~\bibnamefont{Bl\"ugel}}, \bibnamefont{and}
  \bibinfo{author}{\bibfnamefont{J.~B.} \bibnamefont{Staunton}},
  \bibinfo{journal}{Phys. Rev. Lett.} \textbf{\bibinfo{volume}{128}},
  \bibinfo{pages}{157206} (\bibinfo{year}{2022}).

\bibitem[{\citenamefont{Nomoto and Arita}(2023)}]{nomoto2023ab}
\bibinfo{author}{\bibfnamefont{T.}~\bibnamefont{Nomoto}} \bibnamefont{and}
  \bibinfo{author}{\bibfnamefont{R.}~\bibnamefont{Arita}}, \bibinfo{journal}{J.
  Appl. Phys.} \textbf{\bibinfo{volume}{133}} (\bibinfo{year}{2023}).

\bibitem[{\citenamefont{Muto and Mochizuki}(2023)}]{muto2023theory}
\bibinfo{author}{\bibfnamefont{T.}~\bibnamefont{Muto}} \bibnamefont{and}
  \bibinfo{author}{\bibfnamefont{M.}~\bibnamefont{Mochizuki}},
  \bibinfo{journal}{arXiv:2305.00751}  (\bibinfo{year}{2023}).

\bibitem[{\citenamefont{Eto and Mochizuki}(2021)}]{Eto_PhysRevB.104.104425}
\bibinfo{author}{\bibfnamefont{R.}~\bibnamefont{Eto}} \bibnamefont{and}
  \bibinfo{author}{\bibfnamefont{M.}~\bibnamefont{Mochizuki}},
  \bibinfo{journal}{Phys. Rev. B} \textbf{\bibinfo{volume}{104}},
  \bibinfo{pages}{104425} (\bibinfo{year}{2021}).

\bibitem[{\citenamefont{Heinze et~al.}(2011)\citenamefont{Heinze, von Bergmann,
  Menzel, Brede, Kubetzka, Wiesendanger, Bihlmayer, and
  Bl{\"u}gel}}]{heinze2011spontaneous}
\bibinfo{author}{\bibfnamefont{S.}~\bibnamefont{Heinze}},
  \bibinfo{author}{\bibfnamefont{K.}~\bibnamefont{von Bergmann}},
  \bibinfo{author}{\bibfnamefont{M.}~\bibnamefont{Menzel}},
  \bibinfo{author}{\bibfnamefont{J.}~\bibnamefont{Brede}},
  \bibinfo{author}{\bibfnamefont{A.}~\bibnamefont{Kubetzka}},
  \bibinfo{author}{\bibfnamefont{R.}~\bibnamefont{Wiesendanger}},
  \bibinfo{author}{\bibfnamefont{G.}~\bibnamefont{Bihlmayer}},
  \bibnamefont{and}
  \bibinfo{author}{\bibfnamefont{S.}~\bibnamefont{Bl{\"u}gel}},
  \bibinfo{journal}{Nat. Phys.} \textbf{\bibinfo{volume}{7}},
  \bibinfo{pages}{713} (\bibinfo{year}{2011}).

\bibitem[{\citenamefont{Yambe and Hayami}(2023)}]{Yambe_PhysRevB.107.014417}
\bibinfo{author}{\bibfnamefont{R.}~\bibnamefont{Yambe}} \bibnamefont{and}
  \bibinfo{author}{\bibfnamefont{S.}~\bibnamefont{Hayami}},
  \bibinfo{journal}{Phys. Rev. B} \textbf{\bibinfo{volume}{107}},
  \bibinfo{pages}{014417} (\bibinfo{year}{2023}).

\bibitem[{\citenamefont{Hamamoto et~al.}(2015)\citenamefont{Hamamoto, Ezawa,
  and Nagaosa}}]{Hamamoto_PhysRevB.92.115417}
\bibinfo{author}{\bibfnamefont{K.}~\bibnamefont{Hamamoto}},
  \bibinfo{author}{\bibfnamefont{M.}~\bibnamefont{Ezawa}}, \bibnamefont{and}
  \bibinfo{author}{\bibfnamefont{N.}~\bibnamefont{Nagaosa}},
  \bibinfo{journal}{Phys. Rev. B} \textbf{\bibinfo{volume}{92}},
  \bibinfo{pages}{115417} (\bibinfo{year}{2015}).

\bibitem[{\citenamefont{G\"obel
  et~al.}(2017{\natexlab{a}})\citenamefont{G\"obel, Mook, Henk, and
  Mertig}}]{Gobel_PhysRevB.95.094413}
\bibinfo{author}{\bibfnamefont{B.}~\bibnamefont{G\"obel}},
  \bibinfo{author}{\bibfnamefont{A.}~\bibnamefont{Mook}},
  \bibinfo{author}{\bibfnamefont{J.}~\bibnamefont{Henk}}, \bibnamefont{and}
  \bibinfo{author}{\bibfnamefont{I.}~\bibnamefont{Mertig}},
  \bibinfo{journal}{Phys. Rev. B} \textbf{\bibinfo{volume}{95}},
  \bibinfo{pages}{094413} (\bibinfo{year}{2017}{\natexlab{a}}).

\bibitem[{\citenamefont{G\"obel
  et~al.}(2017{\natexlab{b}})\citenamefont{G\"obel, Mook, Henk, and
  Mertig}}]{Gobel_PhysRevB.96.060406}
\bibinfo{author}{\bibfnamefont{B.}~\bibnamefont{G\"obel}},
  \bibinfo{author}{\bibfnamefont{A.}~\bibnamefont{Mook}},
  \bibinfo{author}{\bibfnamefont{J.}~\bibnamefont{Henk}}, \bibnamefont{and}
  \bibinfo{author}{\bibfnamefont{I.}~\bibnamefont{Mertig}},
  \bibinfo{journal}{Phys. Rev. B} \textbf{\bibinfo{volume}{96}},
  \bibinfo{pages}{060406} (\bibinfo{year}{2017}{\natexlab{b}}).

\bibitem[{\citenamefont{Agterberg and
  Yunoki}(2000)}]{Agterberg_PhysRevB.62.13816}
\bibinfo{author}{\bibfnamefont{D.~F.} \bibnamefont{Agterberg}}
  \bibnamefont{and} \bibinfo{author}{\bibfnamefont{S.}~\bibnamefont{Yunoki}},
  \bibinfo{journal}{Phys. Rev. B} \textbf{\bibinfo{volume}{62}},
  \bibinfo{pages}{13816} (\bibinfo{year}{2000}).

\bibitem[{\citenamefont{Gastiasoro and
  Andersen}(2015)}]{Gastiasoro_PhysRevB.92.140506}
\bibinfo{author}{\bibfnamefont{M.~N.} \bibnamefont{Gastiasoro}}
  \bibnamefont{and} \bibinfo{author}{\bibfnamefont{B.~M.}
  \bibnamefont{Andersen}}, \bibinfo{journal}{Phys. Rev. B}
  \textbf{\bibinfo{volume}{92}}, \bibinfo{pages}{140506}
  (\bibinfo{year}{2015}).

\bibitem[{\citenamefont{Allred et~al.}(2016)\citenamefont{Allred, Taddei,
  Bugaris, Krogstad, Lapidus, Chung, Claus, Kanatzidis, Brown, Kang
  et~al.}}]{allred2016double}
\bibinfo{author}{\bibfnamefont{J.}~\bibnamefont{Allred}},
  \bibinfo{author}{\bibfnamefont{K.}~\bibnamefont{Taddei}},
  \bibinfo{author}{\bibfnamefont{D.}~\bibnamefont{Bugaris}},
  \bibinfo{author}{\bibfnamefont{M.}~\bibnamefont{Krogstad}},
  \bibinfo{author}{\bibfnamefont{S.}~\bibnamefont{Lapidus}},
  \bibinfo{author}{\bibfnamefont{D.}~\bibnamefont{Chung}},
  \bibinfo{author}{\bibfnamefont{H.}~\bibnamefont{Claus}},
  \bibinfo{author}{\bibfnamefont{M.}~\bibnamefont{Kanatzidis}},
  \bibinfo{author}{\bibfnamefont{D.}~\bibnamefont{Brown}},
  \bibinfo{author}{\bibfnamefont{J.}~\bibnamefont{Kang}}, \bibnamefont{et~al.},
  \bibinfo{journal}{Nat. Phys.} \textbf{\bibinfo{volume}{12}},
  \bibinfo{pages}{493} (\bibinfo{year}{2016}).

\bibitem[{\citenamefont{Hayami and
  Motome}(2018)}]{Hayami_PhysRevLett.121.137202}
\bibinfo{author}{\bibfnamefont{S.}~\bibnamefont{Hayami}} \bibnamefont{and}
  \bibinfo{author}{\bibfnamefont{Y.}~\bibnamefont{Motome}},
  \bibinfo{journal}{Phys. Rev. Lett.} \textbf{\bibinfo{volume}{121}},
  \bibinfo{pages}{137202} (\bibinfo{year}{2018}).

\bibitem[{\citenamefont{Wang et~al.}(2021)\citenamefont{Wang, Su, Lin, and
  Batista}}]{Wang_PhysRevB.103.104408}
\bibinfo{author}{\bibfnamefont{Z.}~\bibnamefont{Wang}},
  \bibinfo{author}{\bibfnamefont{Y.}~\bibnamefont{Su}},
  \bibinfo{author}{\bibfnamefont{S.-Z.} \bibnamefont{Lin}}, \bibnamefont{and}
  \bibinfo{author}{\bibfnamefont{C.~D.} \bibnamefont{Batista}},
  \bibinfo{journal}{Phys. Rev. B} \textbf{\bibinfo{volume}{103}},
  \bibinfo{pages}{104408} (\bibinfo{year}{2021}).

\bibitem[{\citenamefont{Hayami}(2022{\natexlab{b}})}]{Hayami_PhysRevB.105.174437}
\bibinfo{author}{\bibfnamefont{S.}~\bibnamefont{Hayami}},
  \bibinfo{journal}{Phys. Rev. B} \textbf{\bibinfo{volume}{105}},
  \bibinfo{pages}{174437} (\bibinfo{year}{2022}{\natexlab{b}}).

\bibitem[{\citenamefont{Christensen et~al.}(2018)\citenamefont{Christensen,
  Andersen, and Kotetes}}]{Christensen_PhysRevX.8.041022}
\bibinfo{author}{\bibfnamefont{M.~H.} \bibnamefont{Christensen}},
  \bibinfo{author}{\bibfnamefont{B.~M.} \bibnamefont{Andersen}},
  \bibnamefont{and} \bibinfo{author}{\bibfnamefont{P.}~\bibnamefont{Kotetes}},
  \bibinfo{journal}{Phys. Rev. X} \textbf{\bibinfo{volume}{8}},
  \bibinfo{pages}{041022} (\bibinfo{year}{2018}).

\bibitem[{\citenamefont{Hayami and
  Motome}(2021{\natexlab{c}})}]{Hayami_PhysRevB.103.024439}
\bibinfo{author}{\bibfnamefont{S.}~\bibnamefont{Hayami}} \bibnamefont{and}
  \bibinfo{author}{\bibfnamefont{Y.}~\bibnamefont{Motome}},
  \bibinfo{journal}{Phys. Rev. B} \textbf{\bibinfo{volume}{103}},
  \bibinfo{pages}{024439} (\bibinfo{year}{2021}{\natexlab{c}}).

\bibitem[{\citenamefont{Hayami}(2022{\natexlab{c}})}]{hayami2022multiple}
\bibinfo{author}{\bibfnamefont{S.}~\bibnamefont{Hayami}}, \bibinfo{journal}{J.
  Phys. Soc. Jpn.} \textbf{\bibinfo{volume}{91}}, \bibinfo{pages}{023705}
  (\bibinfo{year}{2022}{\natexlab{c}}).

\bibitem[{\citenamefont{Hayami}(2022{\natexlab{d}})}]{hayami2022square}
\bibinfo{author}{\bibfnamefont{S.}~\bibnamefont{Hayami}}, \bibinfo{journal}{J.
  Phys.: Condens. Matter} \textbf{\bibinfo{volume}{34}},
  \bibinfo{pages}{365802} (\bibinfo{year}{2022}{\natexlab{d}}).

\bibitem[{\citenamefont{Huang and
  Kotetes}(2023)}]{Huang_PhysRevResearch.5.013125}
\bibinfo{author}{\bibfnamefont{Y.-P.} \bibnamefont{Huang}} \bibnamefont{and}
  \bibinfo{author}{\bibfnamefont{P.}~\bibnamefont{Kotetes}},
  \bibinfo{journal}{Phys. Rev. Res.} \textbf{\bibinfo{volume}{5}},
  \bibinfo{pages}{013125} (\bibinfo{year}{2023}).

\bibitem[{\citenamefont{Khanh et~al.}(2020)\citenamefont{Khanh, Nakajima, Yu,
  Gao, Shibata, Hirschberger, Yamasaki, Sagayama, Nakao, Peng
  et~al.}}]{khanh2020nanometric}
\bibinfo{author}{\bibfnamefont{N.~D.} \bibnamefont{Khanh}},
  \bibinfo{author}{\bibfnamefont{T.}~\bibnamefont{Nakajima}},
  \bibinfo{author}{\bibfnamefont{X.}~\bibnamefont{Yu}},
  \bibinfo{author}{\bibfnamefont{S.}~\bibnamefont{Gao}},
  \bibinfo{author}{\bibfnamefont{K.}~\bibnamefont{Shibata}},
  \bibinfo{author}{\bibfnamefont{M.}~\bibnamefont{Hirschberger}},
  \bibinfo{author}{\bibfnamefont{Y.}~\bibnamefont{Yamasaki}},
  \bibinfo{author}{\bibfnamefont{H.}~\bibnamefont{Sagayama}},
  \bibinfo{author}{\bibfnamefont{H.}~\bibnamefont{Nakao}},
  \bibinfo{author}{\bibfnamefont{L.}~\bibnamefont{Peng}}, \bibnamefont{et~al.},
  \bibinfo{journal}{Nat. Nanotechnol.} \textbf{\bibinfo{volume}{15}},
  \bibinfo{pages}{444} (\bibinfo{year}{2020}).

\bibitem[{\citenamefont{Yasui et~al.}(2020)\citenamefont{Yasui, Butler, Khanh,
  Hayami, Nomoto, Hanaguri, Motome, Arita, h.~Arima, Tokura
  et~al.}}]{Yasui2020imaging}
\bibinfo{author}{\bibfnamefont{Y.}~\bibnamefont{Yasui}},
  \bibinfo{author}{\bibfnamefont{C.~J.} \bibnamefont{Butler}},
  \bibinfo{author}{\bibfnamefont{N.~D.} \bibnamefont{Khanh}},
  \bibinfo{author}{\bibfnamefont{S.}~\bibnamefont{Hayami}},
  \bibinfo{author}{\bibfnamefont{T.}~\bibnamefont{Nomoto}},
  \bibinfo{author}{\bibfnamefont{T.}~\bibnamefont{Hanaguri}},
  \bibinfo{author}{\bibfnamefont{Y.}~\bibnamefont{Motome}},
  \bibinfo{author}{\bibfnamefont{R.}~\bibnamefont{Arita}},
  \bibinfo{author}{\bibfnamefont{T.}~\bibnamefont{h.~Arima}},
  \bibinfo{author}{\bibfnamefont{Y.}~\bibnamefont{Tokura}},
  \bibnamefont{et~al.}, \bibinfo{journal}{Nat. Commun.}
  \textbf{\bibinfo{volume}{11}}, \bibinfo{pages}{5925} (\bibinfo{year}{2020}).

\bibitem[{\citenamefont{Nomoto et~al.}(2020)\citenamefont{Nomoto, Koretsune,
  and Arita}}]{Nomoto_PhysRevLett.125.117204}
\bibinfo{author}{\bibfnamefont{T.}~\bibnamefont{Nomoto}},
  \bibinfo{author}{\bibfnamefont{T.}~\bibnamefont{Koretsune}},
  \bibnamefont{and} \bibinfo{author}{\bibfnamefont{R.}~\bibnamefont{Arita}},
  \bibinfo{journal}{Phys. Rev. Lett.} \textbf{\bibinfo{volume}{125}},
  \bibinfo{pages}{117204} (\bibinfo{year}{2020}).

\bibitem[{\citenamefont{Khanh et~al.}(2022)\citenamefont{Khanh, Nakajima,
  Hayami, Gao, Yamasaki, Sagayama, Nakao, Takagi, Motome, Tokura
  et~al.}}]{khanh2022zoology}
\bibinfo{author}{\bibfnamefont{N.~D.} \bibnamefont{Khanh}},
  \bibinfo{author}{\bibfnamefont{T.}~\bibnamefont{Nakajima}},
  \bibinfo{author}{\bibfnamefont{S.}~\bibnamefont{Hayami}},
  \bibinfo{author}{\bibfnamefont{S.}~\bibnamefont{Gao}},
  \bibinfo{author}{\bibfnamefont{Y.}~\bibnamefont{Yamasaki}},
  \bibinfo{author}{\bibfnamefont{H.}~\bibnamefont{Sagayama}},
  \bibinfo{author}{\bibfnamefont{H.}~\bibnamefont{Nakao}},
  \bibinfo{author}{\bibfnamefont{R.}~\bibnamefont{Takagi}},
  \bibinfo{author}{\bibfnamefont{Y.}~\bibnamefont{Motome}},
  \bibinfo{author}{\bibfnamefont{Y.}~\bibnamefont{Tokura}},
  \bibnamefont{et~al.}, \bibinfo{journal}{Adv. Sci.}
  \textbf{\bibinfo{volume}{9}}, \bibinfo{pages}{2105452}
  (\bibinfo{year}{2022}).

\bibitem[{\citenamefont{Matsuyama et~al.}(2023)\citenamefont{Matsuyama, Nomura,
  Imajo, Nomoto, Arita, Sudo, Kimata, Khanh, Takagi, Tokura
  et~al.}}]{Matsuyama_PhysRevB.107.104421}
\bibinfo{author}{\bibfnamefont{N.}~\bibnamefont{Matsuyama}},
  \bibinfo{author}{\bibfnamefont{T.}~\bibnamefont{Nomura}},
  \bibinfo{author}{\bibfnamefont{S.}~\bibnamefont{Imajo}},
  \bibinfo{author}{\bibfnamefont{T.}~\bibnamefont{Nomoto}},
  \bibinfo{author}{\bibfnamefont{R.}~\bibnamefont{Arita}},
  \bibinfo{author}{\bibfnamefont{K.}~\bibnamefont{Sudo}},
  \bibinfo{author}{\bibfnamefont{M.}~\bibnamefont{Kimata}},
  \bibinfo{author}{\bibfnamefont{N.~D.} \bibnamefont{Khanh}},
  \bibinfo{author}{\bibfnamefont{R.}~\bibnamefont{Takagi}},
  \bibinfo{author}{\bibfnamefont{Y.}~\bibnamefont{Tokura}},
  \bibnamefont{et~al.}, \bibinfo{journal}{Phys. Rev. B}
  \textbf{\bibinfo{volume}{107}}, \bibinfo{pages}{104421}
  (\bibinfo{year}{2023}).

\bibitem[{\citenamefont{Wood et~al.}(2023)\citenamefont{Wood, Khalyavin, Mayoh,
  Bouaziz, Hall, Holt, Orlandi, Manuel, Bl\"ugel, Staunton
  et~al.}}]{Wood_PhysRevB.107.L180402}
\bibinfo{author}{\bibfnamefont{G.~D.~A.} \bibnamefont{Wood}},
  \bibinfo{author}{\bibfnamefont{D.~D.} \bibnamefont{Khalyavin}},
  \bibinfo{author}{\bibfnamefont{D.~A.} \bibnamefont{Mayoh}},
  \bibinfo{author}{\bibfnamefont{J.}~\bibnamefont{Bouaziz}},
  \bibinfo{author}{\bibfnamefont{A.~E.} \bibnamefont{Hall}},
  \bibinfo{author}{\bibfnamefont{S.~J.~R.} \bibnamefont{Holt}},
  \bibinfo{author}{\bibfnamefont{F.}~\bibnamefont{Orlandi}},
  \bibinfo{author}{\bibfnamefont{P.}~\bibnamefont{Manuel}},
  \bibinfo{author}{\bibfnamefont{S.}~\bibnamefont{Bl\"ugel}},
  \bibinfo{author}{\bibfnamefont{J.~B.} \bibnamefont{Staunton}},
  \bibnamefont{et~al.}, \bibinfo{journal}{Phys. Rev. B}
  \textbf{\bibinfo{volume}{107}}, \bibinfo{pages}{L180402}
  (\bibinfo{year}{2023}).

\bibitem[{\citenamefont{Shang et~al.}(2021)\citenamefont{Shang, Xu, Gawryluk,
  Ma, Shiroka, Shi, and Pomjakushina}}]{Shang_PhysRevB.103.L020405}
\bibinfo{author}{\bibfnamefont{T.}~\bibnamefont{Shang}},
  \bibinfo{author}{\bibfnamefont{Y.}~\bibnamefont{Xu}},
  \bibinfo{author}{\bibfnamefont{D.~J.} \bibnamefont{Gawryluk}},
  \bibinfo{author}{\bibfnamefont{J.~Z.} \bibnamefont{Ma}},
  \bibinfo{author}{\bibfnamefont{T.}~\bibnamefont{Shiroka}},
  \bibinfo{author}{\bibfnamefont{M.}~\bibnamefont{Shi}}, \bibnamefont{and}
  \bibinfo{author}{\bibfnamefont{E.}~\bibnamefont{Pomjakushina}},
  \bibinfo{journal}{Phys. Rev. B} \textbf{\bibinfo{volume}{103}},
  \bibinfo{pages}{L020405} (\bibinfo{year}{2021}).

\bibitem[{\citenamefont{Shimomura et~al.}(2019)\citenamefont{Shimomura, Murao,
  Tsutsui, Nakao, Nakamura, Hedo, Nakama, and
  {\=O}nuki}}]{shimomura2019lattice}
\bibinfo{author}{\bibfnamefont{S.}~\bibnamefont{Shimomura}},
  \bibinfo{author}{\bibfnamefont{H.}~\bibnamefont{Murao}},
  \bibinfo{author}{\bibfnamefont{S.}~\bibnamefont{Tsutsui}},
  \bibinfo{author}{\bibfnamefont{H.}~\bibnamefont{Nakao}},
  \bibinfo{author}{\bibfnamefont{A.}~\bibnamefont{Nakamura}},
  \bibinfo{author}{\bibfnamefont{M.}~\bibnamefont{Hedo}},
  \bibinfo{author}{\bibfnamefont{T.}~\bibnamefont{Nakama}}, \bibnamefont{and}
  \bibinfo{author}{\bibfnamefont{Y.}~\bibnamefont{{\=O}nuki}},
  \bibinfo{journal}{J. Phys. Soc. Jpn.} \textbf{\bibinfo{volume}{88}},
  \bibinfo{pages}{014602} (\bibinfo{year}{2019}).

\bibitem[{\citenamefont{Kaneko et~al.}(2021)\citenamefont{Kaneko, Kawasaki,
  Nakamura, Munakata, Nakao, Hanashima, Kiyanagi, Ohhara, Hedo, Nakama
  et~al.}}]{kaneko2021charge}
\bibinfo{author}{\bibfnamefont{K.}~\bibnamefont{Kaneko}},
  \bibinfo{author}{\bibfnamefont{T.}~\bibnamefont{Kawasaki}},
  \bibinfo{author}{\bibfnamefont{A.}~\bibnamefont{Nakamura}},
  \bibinfo{author}{\bibfnamefont{K.}~\bibnamefont{Munakata}},
  \bibinfo{author}{\bibfnamefont{A.}~\bibnamefont{Nakao}},
  \bibinfo{author}{\bibfnamefont{T.}~\bibnamefont{Hanashima}},
  \bibinfo{author}{\bibfnamefont{R.}~\bibnamefont{Kiyanagi}},
  \bibinfo{author}{\bibfnamefont{T.}~\bibnamefont{Ohhara}},
  \bibinfo{author}{\bibfnamefont{M.}~\bibnamefont{Hedo}},
  \bibinfo{author}{\bibfnamefont{T.}~\bibnamefont{Nakama}},
  \bibnamefont{et~al.}, \bibinfo{journal}{J. Phys. Soc. Jpn.}
  \textbf{\bibinfo{volume}{90}}, \bibinfo{pages}{064704}
  (\bibinfo{year}{2021}).

\bibitem[{\citenamefont{Zhu et~al.}(2022)\citenamefont{Zhu, Zhang, Gawryluk,
  Zhen, Yu, Ju, Xie, Jiang, Cheng, Xu et~al.}}]{Zhu_PhysRevB.105.014423}
\bibinfo{author}{\bibfnamefont{X.~Y.} \bibnamefont{Zhu}},
  \bibinfo{author}{\bibfnamefont{H.}~\bibnamefont{Zhang}},
  \bibinfo{author}{\bibfnamefont{D.~J.} \bibnamefont{Gawryluk}},
  \bibinfo{author}{\bibfnamefont{Z.~X.} \bibnamefont{Zhen}},
  \bibinfo{author}{\bibfnamefont{B.~C.} \bibnamefont{Yu}},
  \bibinfo{author}{\bibfnamefont{S.~L.} \bibnamefont{Ju}},
  \bibinfo{author}{\bibfnamefont{W.}~\bibnamefont{Xie}},
  \bibinfo{author}{\bibfnamefont{D.~M.} \bibnamefont{Jiang}},
  \bibinfo{author}{\bibfnamefont{W.~J.} \bibnamefont{Cheng}},
  \bibinfo{author}{\bibfnamefont{Y.}~\bibnamefont{Xu}}, \bibnamefont{et~al.},
  \bibinfo{journal}{Phys. Rev. B} \textbf{\bibinfo{volume}{105}},
  \bibinfo{pages}{014423} (\bibinfo{year}{2022}).

\bibitem[{\citenamefont{Takagi et~al.}(2022)\citenamefont{Takagi, Matsuyama,
  Ukleev, Yu, White, Francoual, Mardegan, Hayami, Saito, Kaneko
  et~al.}}]{takagi2022square}
\bibinfo{author}{\bibfnamefont{R.}~\bibnamefont{Takagi}},
  \bibinfo{author}{\bibfnamefont{N.}~\bibnamefont{Matsuyama}},
  \bibinfo{author}{\bibfnamefont{V.}~\bibnamefont{Ukleev}},
  \bibinfo{author}{\bibfnamefont{L.}~\bibnamefont{Yu}},
  \bibinfo{author}{\bibfnamefont{J.~S.} \bibnamefont{White}},
  \bibinfo{author}{\bibfnamefont{S.}~\bibnamefont{Francoual}},
  \bibinfo{author}{\bibfnamefont{J.~R.~L.} \bibnamefont{Mardegan}},
  \bibinfo{author}{\bibfnamefont{S.}~\bibnamefont{Hayami}},
  \bibinfo{author}{\bibfnamefont{H.}~\bibnamefont{Saito}},
  \bibinfo{author}{\bibfnamefont{K.}~\bibnamefont{Kaneko}},
  \bibnamefont{et~al.}, \bibinfo{journal}{Nat. Commun.}
  \textbf{\bibinfo{volume}{13}}, \bibinfo{pages}{1472} (\bibinfo{year}{2022}).

\bibitem[{\citenamefont{Meier et~al.}(2022)\citenamefont{Meier, Torres,
  Hermann, Zhao, Lavina, Sales, and May}}]{Meier_PhysRevB.106.094421}
\bibinfo{author}{\bibfnamefont{W.~R.} \bibnamefont{Meier}},
  \bibinfo{author}{\bibfnamefont{J.~R.} \bibnamefont{Torres}},
  \bibinfo{author}{\bibfnamefont{R.~P.} \bibnamefont{Hermann}},
  \bibinfo{author}{\bibfnamefont{J.}~\bibnamefont{Zhao}},
  \bibinfo{author}{\bibfnamefont{B.}~\bibnamefont{Lavina}},
  \bibinfo{author}{\bibfnamefont{B.~C.} \bibnamefont{Sales}}, \bibnamefont{and}
  \bibinfo{author}{\bibfnamefont{A.~F.} \bibnamefont{May}},
  \bibinfo{journal}{Phys. Rev. B} \textbf{\bibinfo{volume}{106}},
  \bibinfo{pages}{094421} (\bibinfo{year}{2022}).

\bibitem[{\citenamefont{Gen et~al.}(2023)\citenamefont{Gen, Takagi, Watanabe,
  Kitou, Sagayama, Matsuyama, Kohama, Ikeda, \ifmmode~\bar{O}\else
  \={O}\fi{}nuki, Kurumaji et~al.}}]{Gen_PhysRevB.107.L020410}
\bibinfo{author}{\bibfnamefont{M.}~\bibnamefont{Gen}},
  \bibinfo{author}{\bibfnamefont{R.}~\bibnamefont{Takagi}},
  \bibinfo{author}{\bibfnamefont{Y.}~\bibnamefont{Watanabe}},
  \bibinfo{author}{\bibfnamefont{S.}~\bibnamefont{Kitou}},
  \bibinfo{author}{\bibfnamefont{H.}~\bibnamefont{Sagayama}},
  \bibinfo{author}{\bibfnamefont{N.}~\bibnamefont{Matsuyama}},
  \bibinfo{author}{\bibfnamefont{Y.}~\bibnamefont{Kohama}},
  \bibinfo{author}{\bibfnamefont{A.}~\bibnamefont{Ikeda}},
  \bibinfo{author}{\bibfnamefont{Y.}~\bibnamefont{\ifmmode~\bar{O}\else
  \={O}\fi{}nuki}}, \bibinfo{author}{\bibfnamefont{T.}~\bibnamefont{Kurumaji}},
  \bibnamefont{et~al.}, \bibinfo{journal}{Phys. Rev. B}
  \textbf{\bibinfo{volume}{107}}, \bibinfo{pages}{L020410}
  (\bibinfo{year}{2023}).

\bibitem[{\citenamefont{Hayami}(2023)}]{hayami2023orthorhombic}
\bibinfo{author}{\bibfnamefont{S.}~\bibnamefont{Hayami}}, \bibinfo{journal}{J.
  Phys.: Mater.} \textbf{\bibinfo{volume}{6}}, \bibinfo{pages}{014006}
  (\bibinfo{year}{2023}).

\bibitem[{\citenamefont{Hayami and Kato}(2023)}]{hayami2023widely}
\bibinfo{author}{\bibfnamefont{S.}~\bibnamefont{Hayami}} \bibnamefont{and}
  \bibinfo{author}{\bibfnamefont{Y.}~\bibnamefont{Kato}}, \bibinfo{journal}{J.
  Magn. Magn. Mater.} \textbf{\bibinfo{volume}{571}}, \bibinfo{pages}{170547}
  (\bibinfo{year}{2023}).

\bibitem[{\citenamefont{Solenov et~al.}(2012)\citenamefont{Solenov, Mozyrsky,
  and Martin}}]{Solenov_PhysRevLett.108.096403}
\bibinfo{author}{\bibfnamefont{D.}~\bibnamefont{Solenov}},
  \bibinfo{author}{\bibfnamefont{D.}~\bibnamefont{Mozyrsky}}, \bibnamefont{and}
  \bibinfo{author}{\bibfnamefont{I.}~\bibnamefont{Martin}},
  \bibinfo{journal}{Phys. Rev. Lett.} \textbf{\bibinfo{volume}{108}},
  \bibinfo{pages}{096403} (\bibinfo{year}{2012}).

\bibitem[{\citenamefont{Ozawa et~al.}(2016)\citenamefont{Ozawa, Hayami, Barros,
  Chern, Motome, and Batista}}]{Ozawa_doi:10.7566/JPSJ.85.103703}
\bibinfo{author}{\bibfnamefont{R.}~\bibnamefont{Ozawa}},
  \bibinfo{author}{\bibfnamefont{S.}~\bibnamefont{Hayami}},
  \bibinfo{author}{\bibfnamefont{K.}~\bibnamefont{Barros}},
  \bibinfo{author}{\bibfnamefont{G.-W.} \bibnamefont{Chern}},
  \bibinfo{author}{\bibfnamefont{Y.}~\bibnamefont{Motome}}, \bibnamefont{and}
  \bibinfo{author}{\bibfnamefont{C.~D.} \bibnamefont{Batista}},
  \bibinfo{journal}{J. Phys. Soc. Jpn.} \textbf{\bibinfo{volume}{85}},
  \bibinfo{pages}{103703} (\bibinfo{year}{2016}).

\bibitem[{\citenamefont{Hayami}(2020)}]{hayami2020multiple}
\bibinfo{author}{\bibfnamefont{S.}~\bibnamefont{Hayami}}, \bibinfo{journal}{J.
  Magn. Magn. Mater.} \textbf{\bibinfo{volume}{513}}, \bibinfo{pages}{167181}
  (\bibinfo{year}{2020}).

\bibitem[{\citenamefont{Hayami and
  Yambe}(2020)}]{Hayami_doi:10.7566/JPSJ.89.103702}
\bibinfo{author}{\bibfnamefont{S.}~\bibnamefont{Hayami}} \bibnamefont{and}
  \bibinfo{author}{\bibfnamefont{R.}~\bibnamefont{Yambe}}, \bibinfo{journal}{J.
  Phys. Soc. Jpn.} \textbf{\bibinfo{volume}{89}}, \bibinfo{pages}{103702}
  (\bibinfo{year}{2020}).

\bibitem[{\citenamefont{Hayami and Yambe}(2022)}]{Hayami_PhysRevB.105.104428}
\bibinfo{author}{\bibfnamefont{S.}~\bibnamefont{Hayami}} \bibnamefont{and}
  \bibinfo{author}{\bibfnamefont{R.}~\bibnamefont{Yambe}},
  \bibinfo{journal}{Phys. Rev. B} \textbf{\bibinfo{volume}{105}},
  \bibinfo{pages}{104428} (\bibinfo{year}{2022}).

\bibitem[{\citenamefont{Fukui et~al.}(2005)\citenamefont{Fukui, Hatsugai, and
  Suzuki}}]{fukui2005chern}
\bibinfo{author}{\bibfnamefont{T.}~\bibnamefont{Fukui}},
  \bibinfo{author}{\bibfnamefont{Y.}~\bibnamefont{Hatsugai}}, \bibnamefont{and}
  \bibinfo{author}{\bibfnamefont{H.}~\bibnamefont{Suzuki}},
  \bibinfo{journal}{J. Phys. Soc. Jpn.} \textbf{\bibinfo{volume}{74}},
  \bibinfo{pages}{1674} (\bibinfo{year}{2005}).

\bibitem[{\citenamefont{Tatara and Kawamura}(2002)}]{tatara2002chirality}
\bibinfo{author}{\bibfnamefont{G.}~\bibnamefont{Tatara}} \bibnamefont{and}
  \bibinfo{author}{\bibfnamefont{H.}~\bibnamefont{Kawamura}},
  \bibinfo{journal}{J. Phys. Soc. Jpn.} \textbf{\bibinfo{volume}{71}},
  \bibinfo{pages}{2613} (\bibinfo{year}{2002}).

\bibitem[{\citenamefont{Nakazawa and Kohno}(2014)}]{nakazawa2014effects}
\bibinfo{author}{\bibfnamefont{K.}~\bibnamefont{Nakazawa}} \bibnamefont{and}
  \bibinfo{author}{\bibfnamefont{H.}~\bibnamefont{Kohno}}, \bibinfo{journal}{J.
  Phys. Soc. Jpn.} \textbf{\bibinfo{volume}{83}}, \bibinfo{pages}{073707}
  (\bibinfo{year}{2014}).

\bibitem[{\citenamefont{Ishizuka and Nagaosa}(2018)}]{ishizuka2018spin}
\bibinfo{author}{\bibfnamefont{H.}~\bibnamefont{Ishizuka}} \bibnamefont{and}
  \bibinfo{author}{\bibfnamefont{N.}~\bibnamefont{Nagaosa}},
  \bibinfo{journal}{Sci. Adv.} \textbf{\bibinfo{volume}{4}},
  \bibinfo{pages}{eaap9962} (\bibinfo{year}{2018}).

\bibitem[{\citenamefont{Nakazawa et~al.}(2018)\citenamefont{Nakazawa, Bibes,
  and Kohno}}]{nakazawa2018topological}
\bibinfo{author}{\bibfnamefont{K.}~\bibnamefont{Nakazawa}},
  \bibinfo{author}{\bibfnamefont{M.}~\bibnamefont{Bibes}}, \bibnamefont{and}
  \bibinfo{author}{\bibfnamefont{H.}~\bibnamefont{Kohno}}, \bibinfo{journal}{J.
  Phys. Soc. Jpn.} \textbf{\bibinfo{volume}{87}}, \bibinfo{pages}{033705}
  (\bibinfo{year}{2018}).

\bibitem[{\citenamefont{Denisov et~al.}(2018)\citenamefont{Denisov, Rozhansky,
  Averkiev, and L\"ahderanta}}]{Denisov_PhysRevB.98.195439}
\bibinfo{author}{\bibfnamefont{K.~S.} \bibnamefont{Denisov}},
  \bibinfo{author}{\bibfnamefont{I.~V.} \bibnamefont{Rozhansky}},
  \bibinfo{author}{\bibfnamefont{N.~S.} \bibnamefont{Averkiev}},
  \bibnamefont{and}
  \bibinfo{author}{\bibfnamefont{E.}~\bibnamefont{L\"ahderanta}},
  \bibinfo{journal}{Phys. Rev. B} \textbf{\bibinfo{volume}{98}},
  \bibinfo{pages}{195439} (\bibinfo{year}{2018}).

\bibitem[{\citenamefont{Rosales et~al.}(2019)\citenamefont{Rosales,
  Albarrac\'{\i}n, and Pujol}}]{Rosales_PhysRevB.99.035163}
\bibinfo{author}{\bibfnamefont{H.~D.} \bibnamefont{Rosales}},
  \bibinfo{author}{\bibfnamefont{F.~A.~G.} \bibnamefont{Albarrac\'{\i}n}},
  \bibnamefont{and} \bibinfo{author}{\bibfnamefont{P.}~\bibnamefont{Pujol}},
  \bibinfo{journal}{Phys. Rev. B} \textbf{\bibinfo{volume}{99}},
  \bibinfo{pages}{035163} (\bibinfo{year}{2019}).

\bibitem[{\citenamefont{Mohanta et~al.}(2019)\citenamefont{Mohanta, Dagotto,
  and Okamoto}}]{Mohanta_PhysRevB.100.064429}
\bibinfo{author}{\bibfnamefont{N.}~\bibnamefont{Mohanta}},
  \bibinfo{author}{\bibfnamefont{E.}~\bibnamefont{Dagotto}}, \bibnamefont{and}
  \bibinfo{author}{\bibfnamefont{S.}~\bibnamefont{Okamoto}},
  \bibinfo{journal}{Phys. Rev. B} \textbf{\bibinfo{volume}{100}},
  \bibinfo{pages}{064429} (\bibinfo{year}{2019}).

\bibitem[{\citenamefont{Bouaziz et~al.}(2021)\citenamefont{Bouaziz, Ishida,
  Lounis, and Bl\"ugel}}]{Bouaziz_PhysRevLett.126.147203}
\bibinfo{author}{\bibfnamefont{J.}~\bibnamefont{Bouaziz}},
  \bibinfo{author}{\bibfnamefont{H.}~\bibnamefont{Ishida}},
  \bibinfo{author}{\bibfnamefont{S.}~\bibnamefont{Lounis}}, \bibnamefont{and}
  \bibinfo{author}{\bibfnamefont{S.}~\bibnamefont{Bl\"ugel}},
  \bibinfo{journal}{Phys. Rev. Lett.} \textbf{\bibinfo{volume}{126}},
  \bibinfo{pages}{147203} (\bibinfo{year}{2021}).

\bibitem[{\citenamefont{Matsui et~al.}(2021)\citenamefont{Matsui, Nomoto, and
  Arita}}]{Matsui_PhysRevB.104.174432}
\bibinfo{author}{\bibfnamefont{A.}~\bibnamefont{Matsui}},
  \bibinfo{author}{\bibfnamefont{T.}~\bibnamefont{Nomoto}}, \bibnamefont{and}
  \bibinfo{author}{\bibfnamefont{R.}~\bibnamefont{Arita}},
  \bibinfo{journal}{Phys. Rev. B} \textbf{\bibinfo{volume}{104}},
  \bibinfo{pages}{174432} (\bibinfo{year}{2021}).

\bibitem[{\citenamefont{Rashba}(1960)}]{rashba1960properties}
\bibinfo{author}{\bibfnamefont{E.~I.} \bibnamefont{Rashba}},
  \bibinfo{journal}{Sov. Phys. Solid State} \textbf{\bibinfo{volume}{2}},
  \bibinfo{pages}{1109} (\bibinfo{year}{1960}).

\bibitem[{com()}]{comment_antisymmetric_2QMC}
\bibinfo{note}{It is noted that there is a degeneracy of the global rotation
  around the $z$ axis in spin space. Depending on the final spin configuration,
  the chiral-type antisymmetric spin splitting in the form of $k_x
  s^x_{\bm{k}}+k_y s^y_{\bm{k}}$ and anti-type one in the form of $k_x
  s^x_{\bm{k}}- k_y s^y_{\bm{k}}$ and $k_x s^y_{\bm{k}}+k_y s^x_{\bm{k}}$
  emerges.}

\bibitem[{\citenamefont{Hayami}(2022{\natexlab{e}})}]{Hayami_PhysRevB.105.024413}
\bibinfo{author}{\bibfnamefont{S.}~\bibnamefont{Hayami}},
  \bibinfo{journal}{Phys. Rev. B} \textbf{\bibinfo{volume}{105}},
  \bibinfo{pages}{024413} (\bibinfo{year}{2022}{\natexlab{e}}).

\bibitem[{\citenamefont{Hayami et~al.}(2020{\natexlab{a}})\citenamefont{Hayami,
  Yanagi, and Kusunose}}]{Hayami_PhysRevB.101.220403}
\bibinfo{author}{\bibfnamefont{S.}~\bibnamefont{Hayami}},
  \bibinfo{author}{\bibfnamefont{Y.}~\bibnamefont{Yanagi}}, \bibnamefont{and}
  \bibinfo{author}{\bibfnamefont{H.}~\bibnamefont{Kusunose}},
  \bibinfo{journal}{Phys. Rev. B} \textbf{\bibinfo{volume}{101}},
  \bibinfo{pages}{220403(R)} (\bibinfo{year}{2020}{\natexlab{a}}).

\bibitem[{\citenamefont{Hayami et~al.}(2020{\natexlab{b}})\citenamefont{Hayami,
  Yanagi, and Kusunose}}]{Hayami_PhysRevB.102.144441}
\bibinfo{author}{\bibfnamefont{S.}~\bibnamefont{Hayami}},
  \bibinfo{author}{\bibfnamefont{Y.}~\bibnamefont{Yanagi}}, \bibnamefont{and}
  \bibinfo{author}{\bibfnamefont{H.}~\bibnamefont{Kusunose}},
  \bibinfo{journal}{Phys. Rev. B} \textbf{\bibinfo{volume}{102}},
  \bibinfo{pages}{144441} (\bibinfo{year}{2020}{\natexlab{b}}).

\bibitem[{\citenamefont{Hayami}(2022{\natexlab{f}})}]{Hayami_PhysRevB.106.144402}
\bibinfo{author}{\bibfnamefont{S.}~\bibnamefont{Hayami}},
  \bibinfo{journal}{Phys. Rev. B} \textbf{\bibinfo{volume}{106}},
  \bibinfo{pages}{144402} (\bibinfo{year}{2022}{\natexlab{f}}).

\end{thebibliography}
\end{document}